\documentclass[a4paper]{article}
\usepackage[margin=0.5in]{geometry}
\usepackage{graphicx}
\usepackage{graphics}
\usepackage{subfigure}
\usepackage{setspace}
\usepackage{amsmath,amssymb}
\usepackage{xspace}
\usepackage[
	backend=bibtex,
	sorting=none,
	firstinits=true,
	maxbibnames=99,minbibnames=99,
	style=numeric-comp
	]{biblatex}
\bibliography{mine}

\newcommand{\isotope} [2] {\ensuremath{{}^{#2}\!{\mathrm{#1}}}}
\newcommand{\Cone}{\ensuremath{\mathrm{C1}}}
\newcommand{\Ctwo}{\ensuremath{\mathrm{C2}}}
\newcommand{\Tone}{\ensuremath{\mathrm{T_1}}}
\newcommand{\Ttwo}{\ensuremath{\mathrm{T_2}}}
\newcommand{\h}{\ensuremath{\mathrm{H}}}

\newcommand{\ket}[1]{\left|#1\right>}

\def        \eps           {\varepsilon}
\def        \epsn          {\epsilon\xspace}
\def        \sec           {\ensuremath{\mathrm{s}}}
\def        \msec          {\ensuremath{\mathrm{ms}}}
\def        \Hz           {\ensuremath{\mathrm{Hz}}\xspace}
\def        \MHz          {\ensuremath{\mathrm{MHz}}\xspace}
\def	    \RR   	  {\ensuremath{\mathcal{R}}\xspace}
\def\Tonec  {\ensuremath{\mathrm{T_1 (comp) }}\xspace} 
\def\Toner  {\ensuremath{\mathrm{T_1 (reset) }}\xspace} 
\def\Trun  {\ensuremath{\mathrm{T_{run}}}\xspace} 
\def\NRes {\ensuremath{\mathrm{N}}\xspace} 
\def\Twait  {\ensuremath{\mathrm{T_{WAIT}}}\xspace} 
\def	    \RR   {\ensuremath{\mathcal{R}}}

\makeatletter
\renewcommand{\maketitle}{\bgroup\setlength{\parindent}{0pt}
\begin{flushleft}
  \textbf{\LARGE\@title}
  \\~\\
  \@author
\end{flushleft}\egroup
}
\makeatother

\begin{document}
\title{Experimental Heat-Bath Cooling of Spins}
\author{G.~Brassard$^1$, Y.~Elias$^2$, J.~M.~Fernandez$^3$, H.~Gilboa$^4$,  J.~A.~Jones$^5$, T.~Mor$^2$, Y.~Weinstein$^2$ and L.~Xiao$^5$
\\~\\
$^1$ {D{\'e}partement IRO, Universit{\'e} de Montr{\'e}al,
Montr{\'e}al  H3C 3J7, Canada}
\\
$^2$ Department of Computer Science, Technion, Haifa 32000, Israel
\\
$^3$ D{\'e}partement de g{\'e}nie informatique, {\'E}cole Polytechnique de Montr{\'e}al, Montr{\'e}al  H3C 3A7, Canada
\\
$^4$ Department of Chemistry, Technion, Haifa 32000, Israel
\\
$^5$ Centre for Quantum Computation, Clarendon Laboratory,
University of Oxford, Parks Road, Oxford OX13PU, United Kingdom
}
\maketitle
\abstract{
Algorithmic cooling (AC) is a method to purify quantum systems, such as ensembles of nuclear spins, or cold atoms in an optical lattice. 
When applied to spins, AC produces ensembles of highly polarized spins, which enhance the signal strength in nuclear magnetic resonance (NMR).
According to this cooling approach, spin-half nuclei in a constant magnetic field 
are considered as bits, or more precisely, quantum bits,
in a known probability distribution. 
Algorithmic steps on these bits are then translated into 
specially designed  
NMR pulse sequences using common NMR quantum computation tools.
The {\em algorithmic} cooling of spins is achieved by alternately
combining reversible, 
entropy-preserving manipulations (borrowed from data
compression algorithms) with 
\emph{selective reset}, the transfer of entropy from selected spins 
to the environment. In theory, applying algorithmic cooling to sufficiently large spin systems may
produce polarizations far beyond the limits due to conservation of Shannon entropy.
\\
Here, only selective reset steps are performed, 
hence we prefer to call this process ``heat-bath'' cooling, rather
than algorithmic cooling.
We experimentally implement here two consecutive steps of selective reset 
that transfer entropy from two selected spins to the environment.
We performed such cooling experiments with com\-mer\-cia\-lly-available labeled
 molecules, on standard liquid-state NMR spectrometers.
Our experiments yielded polarizations that 
\emph{bypass Shannon's entropy-conservation bound}, 
so that the entire spin-system was cooled. 
This paper was initially submitted in 2005, first to Science and
then to PNAS, and includes additional results from subsequent years (e.g. for resubmission in 2007). The PostScriptum includes more details.
} 
\section{Introduction}~\label{sec:intro}
%
%
The limiting factor
in Nuclear Magnetic Resonance (NMR) spectroscopy and
imaging (MRI) is often the intrinsically low polarization of nuclear spins,
which leads to low signal-to-noise ratio (SNR). Enhancement of the SNR would
permit more rapid data acquisition, enabling more efficient analysis of
 chemicals and visualization of tissues. Alternatively, less material would be
required, leading to reduced toxicity.
Possible methods to
increase the sensitivity of NMR~\cite{EBW87,Slichter90} 
include higher magnetic fields which are currently limited to about
20~tesla, signal averaging which is time consuming and
temperature reduction, which are impractical for various biomedical applications. Another
solution is \emph{effective-cooling};
cooling the spins but not their environment,    
so that their \emph{effective temperature} is lower than the temperature of the 
surrounding heat-bath.
A spin-half particle in a constant and homogeneous 
magnetic field has a steady-state
polarization bias, $\eps$, inversely related
to the temperature, $T$. 
Therefore,
spins with polarization biases above their
equilibrium bias are considered \emph{cool}, relative to the  
temperature of the environment~\cite{Slichter90}. 
Polarization enhancement by a factor of $M$ 
improves the SNR by the
same factor. 
For the case relevant here, $\eps \ll 1$, and then it is
inversely \emph{proportional} to the temperature.

%
%
Effective-cooling of spins, also called ``spin cooling'', 
is a well established concept. Many NMR studies
rely on a variety of methods to transfer polarization among spins
to increase 
signal intensity; 
see~\cite{Slichter90,Sorensen89} and references therein. 
The simplest polarization transfer (PT) method, commonly used in NMR, 
is a transfer of polarization from a more polarized (sensitive) spin to a
significantly less polarized (insensitive) spin, e.g., from
 hydrogen (\isotope{H}{1}) to carbon (\isotope{C}{13}). 
Other (less common) PT methods, 
such as dynamic nuclear polarization, transfer
polarization from electron spins to nuclear spins.
Several enhanced effective-cooling methods have 
succeeded in creating very high polarizations by what one may call the 
``deep-freeze'' approach, in which extreme physical cooling of the relevant
spins 
is involved. These include  
dynamic nuclear polarization 
enhanced with physical cooling and heating~\cite{AFGH+03}, 
\emph{para}-hydrogen in two-spin
systems~\cite{JHBG06,ABCD+04} and hyperpolarized 
xenon~\cite{FSH98,VLSVC01,OS04}.
We briefly describe various effective-cooling methods 
in Appendix~\ref{app:effective-cooling-approaches-in-NMR}.

%
%
Two other approaches, suggested by  
S{\o}rensen (1989) and by
Schulman and Vazirani (1999), are
fundamentally different from the above. They are based on
general unitary transformations~\cite{Sorensen89} and on closely
related data compression methods in closed systems~\cite{SV99}.
Both approaches use reversible entropy-preserving manipulations to cool
specific spins, while heating others.
These approaches are distinct from the simple PT 
and the deep-freeze approaches,
as they allow cooling beyond the capacity of the high-bias spins.
However, those entropy manipulation approaches in 
closed systems~\cite{Sorensen89,SV99}
are limited by Shannon's
entropy-conservation bound~\cite{Ash90,CT91} and S{\o}rensen's unitarity 
bound~\cite{Sorensen89}.
The connection between data compression,
entropy manipulations, and the resulting  
increased polarization bias 
is explained in~\cite{SV99,BMRVV02,FLMR04}; see also 
Appendix~\ref{app:loss-less-compr-to-polarization-compr}.

%
%
A more general approach,
\emph{algorithmic cooling} (AC), 
introduced by Boykin,  
Mor, Roychowdhury, Vatan and Vrijen,
in 2002, makes use of entropy manipulations in
open systems~\cite{BMRVV02}. AC cools beyond 
the reversible schemes~\cite{Sorensen89,SV99} described above.
In particular, when applied to spins, AC can cool beyond Shannon's 
entropy bound  
on closed-system entropy manipulations
(the source coding
theorem~\cite{Ash90,CT91}).

%
%
The entropy manipulation approaches in closed systems~\cite{SV99}
and open systems~\cite{BMRVV02} 
were originally suggested as methods for 
conceptually resolving 
(from a computational complexity point of view) 
the scalability
problem of NMR Quantum Computing (NMRQC).
The open-system method, AC, was later recognized also for its
potential usefulness in NMR spectroscopy~\cite{FLMR04,AAC-pat},
since moderate 
effective-cooling beyond Shannon's entropy bound
is theoretically achievable 
even when AC is applied onto small molecules.

\section{Cooling of Spins Beyond Shannon's Entropy Bound}
%
%
{}From an information theoretic point of view, 
that is, when arbitrary 
closed-system entropy-preserving manipulations are 
allowed~\cite{Sorensen89,SV99}, 
the limitations due to 
Shannon's bound on closed-system entropy manipulations 
can be shown succinctly via a simple
example\footnote{For a different example~\cite{FLMR04} 
see Appendix~\ref{app:loss-less-compr-to-polarization-compr}.}:
Consider our experimental 3-spin system, trichloroethylene,
and recall that we assume polarization biases much smaller than one.
This molecule contains three relevant spins; two labeled carbons   
with nearly identical equilibrium
polarizations, and one hydrogen spin with an equilibrium polarization
that is nearly four-fold higher. For simplicity, we take these polarizations in
this section as $\eps$ for both carbons and $4\eps$ for the hydrogen spin.
The Shannon entropy of the system at
thermal equilibrium is calculated (to leading order in $\eps$)
by summing the entropy of the three spins, $H_{\eps} + H_{\eps} + H_{4\eps}$,
where $H_\eps \triangleq -\frac{1+\eps}{2}\log_2\left(\frac{1+\eps}{2}\right) -
\frac{1-\eps}{2}\log_2\left(\frac{1-\eps}{2}\right) \xrightarrow{\eps\ll 1}
 1-\eps^2/\ln4,$ such that
 $H = 3-(1^2+1^2+4^2)\eps^2/\ln 4 \longrightarrow 3-18\eps^2/\ln4$,
in \emph{bit} units.
The \emph{information content} of the molecule, that is the
difference from maximal entropy, is given in this case by
$3-H,$ such that to very good approximation, in units of $\eps^2/\ln 4,$ the
 initial information content is
\begin{equation}\label{I-initial}
I_{initial}^{approx} \approx 18
\ . \end{equation}
In this paper we are mainly interested in increasing the
value of $I$, thus decreasing the total entropy $(H)$ and cooling the
entire system.
Bypassing Shannon's entropy bound on this 3-spin system
means increasing the information content
of the system above 18.
This could be done,  
for instance, by increasing the polarization of a single spin above the 
value of $\sqrt{18}\eps$ (see the Post-Scriptum for recent experimental results
achieving that goal). 
It could also be done by
reaching biases above 
the value of $\sqrt{6}\eps$ on the three spins, or reaching biases
of $2$ on two spins and a bias above $\sqrt{10}$ on the third. 

%
%
AC was introduced~\cite{BMRVV02} in order 
to increase spin polarization beyond Shannon's entropy bound, 
and, for long molecules, even \emph{far beyond} it (at least in theory).
Notably, thermalization is used beneficially as an integral part
of the cooling scheme, while ordinarily it is perceived
as a major obstacle in quantum computation and NMRQC\@. 
AC employs slow-relaxing spins named \emph{computation spins} and
rapidly relaxing spins named \emph{reset spins}, 
to cool the entire spin-system by pumping
entropy to the environment.
The ratio \RR, between the spin-lattice
relaxation (thermalization) times of the computation spins and the reset spins, must satisfy
$\RR \gg 1$, to permit the application of many cooling steps to
the system, while the computation spins are still quite isolated from the 
environment, namely from the heat-bath. 
In principle, here are the three basic
operations of AC~\cite{BMRVV02,FLMR04}:  
a.---   COMPRESSION. 
Reversible entropy manipulation 
steps (on more than two spins) that redistribute the
   entropy in the system so that some spins become
cooler than the environment, while other
    spins become hotter~\cite{Sorensen89,SV99}.
Saying that certain spins became cooler implies that they became 
more polarized.
b.---   SWAP. Controlled interactions allow specific computation spins to
   adiabatically lose their entropy to reset spins that have 
lower entropy. This means that the polarization is transferred 
from the reset spins onto these computation spins.
c.---   WAIT. The reset spins rapidly thermalize, transferring their entropy to
   the environment, while the computation spins remain colder, so that the
   entire system is cooled.

%
%
This combined set of operations increases spin polarization 
and bypasses Shannon's entropy bound.
AC uses these operations recursively, in order to obtain, 
theoretically, extremely low spin temperatures.
Note that although the algorithms are classical (namely, can be 
defined using only 
classical bits), the implementations via spins make use of the tools
developed in NMRQC 
such as the use of specific quantum gates
(see reviews in~\cite{CLK+00,Jones01}).
In particular,
a universal set of gates (by definition) can be used to compose any algorithm,  
yet a subset might also be sufficient for running a specific algorithm such as
AC\@.

%
%
With sufficiently large relaxation-times ratio 
($\RR \gg 1$) and identical initial bias ($\eps$) for all 
computation and reset spins,  
AC can 
theoretically cool exponentially more than the  
closed-system entropy-preserving 
method:  
In 2004, Fernandez, Lloyd, Mor and Roychowdhury~\cite{FLMR04}
designed cooling algorithms, 
named practicable AC (PAC) that
use simple quantum gates, and that 
can be applied to
\emph{short molecules}; 
As long as the final bias is small\footnote{AC is good  
also for final biases approaching 1~\cite{SMW05}.}, 
$\eps_{final} \ll 1$, 
the application of a simple algorithm to $n-1$ 
computation spins and  
one reset spin ($n$ is odd here, in this example) theoretically  
improves the polarization of a single  
computation spin by 
an exponential factor, yielding 
$ \eps_{final} \approx (3/2)^{(n-1)/2} \eps $,
while Shannon's bound restricts reversible cooling to  
$ \eps_{final} \approx \sqrt{n} \eps$; 
see Appendix~\ref{app:reversible-compression-and-AC}, 
sections~\ref{app:loss-less-compr-to-polarization-compr} 
and~\ref{app:AC-brief-review}.

%
%
AC can (in theory) yield interesting nontrivial 
cooling, 
even on our simple 3-spin system described earlier,  
provided that the (hydrogen) spin with the higher initial polarization of 
$4\eps$ is \emph{also} a reset spin. 
In that case, AC can ideally increase the polarization 
of one carbon spin by a factor of $6$ within very few 
steps,
and 
asymptotically (in the limit of an infinite number
of steps~\cite{Jose-PhD-Thesis}) by a factor of $8$.
In comparison,
PT can at most increase the polarization of one carbon spin 
by a factor of $4$, and the optimal closed-system 
cooling is limited to cooling one spin by a factor of $\sqrt{18}$.
The information content of our 3-spin system 
can (ideally) be increased using very simple AC protocols
to 48 or even 68 within very few steps, 
 and asymptotically to 96.  
For the derivations of those numbers, see Appendix~\ref{app:theoretical-HBC-and-AC-of-TCE}.

%
%
In reality, finite thermalization time ratios  
make cooling beyond Shannon's bound challenging, 
especially if one prefers
to use conventional liquid-state NMR systems  
and commercially-available molecules. 
Realistic AC suffers from small \RR\ 
between the relaxation times ({\Tone}) of relevant computation and reset spins
(${\Tone}^{comp}$ and ${\Tone}^{reset}$, respectively),
and the conditions under which exponential cooling
is still possible are 
not yet fully understood.  
Taking into consideration this gap and others (that are described 
later) 
between theory and experiment, it is
not at all clear whether AC can yield the desired 
advantages even when applied to short molecules.
In particular, for liquid-state NMR, the typical range of R in 
 commercially-available candidate molecules is around 2-20.

\section{Heat-bath Cooling of Spins}
%
%
We apply here a simplified variant of AC, which we 
call ``heat-bath cooling'', to the 3-spin system trichloroethylene, 
shown 
in Figure~\ref{fig:TCE};
we experimentally bypass 
Shannon's entropy bound
with the spin-system of this organic molecule.
Our results, initially appearing in the public domain in 2005~\cite{POTENT}, have encouraged later research 
in this direction,
which could eventually lead to applications already in the near 
future,
as even moderate cooling can be found useful in various 
bio-medical NMR applications. 
This is in contrast with other applications of quantum
computing, such as Shor's factorization algorithm, that probably
cannot become useful in the near future.

%
%
To define heat-bath cooling, let us
look again at the steps that compose algorithmic cooling.
The three steps of AC can be 
combined together (interlaced)
in several ways, yielding slightly modified definitions of 
the building blocks composing AC\@. 
For instance, compression steps can be applied directly on
computational spins and reset spins together to compress entropy onto
one specific reset spin~\cite{FLMR04,Jose-PhD-Thesis,SMW05}. 
Seen this way, AC is composed of (general) reversible 
entropy-preserving manipulations (containing COMPRESSION and SWAP 
steps as special cases) combined with WAIT steps. 

%
%
Alternatively, one may view AC as being composed of
COMPRESSION (or more generally, reversible entropy-preserving manipulations)
and \emph{selective reset} steps, in which  
the entropy is transferred from 
{\em selected} computation spins 
to the heat-bath.
Each such selective-reset step is composed solely of SWAP
and WAIT steps, aimed at cooling a specific
computation spin 
by transferring  
its entropy to a specific (and more polarized) reset spin, and then waiting.
Consequently, that reset spin 
thermalizes and
conveys its excess entropy irreversibly to the heat-bath. 
The refreshed reset spin can then be reused for 
additional selective resets,
e.g. to cool spins that are heated in a subsequent    
polarization compression. 
Each selective reset step thus 
potentially cools the entire spin system.

%
%
When selective-reset steps are 
performed, 
with no compression step(s),
in order to cool computation spins,
we refer to this degenerate
version of AC as ``heat-bath cooling''.  
Here we
experimentally show controlled entropy
extraction from our 
3-spin system via a \emph{dual-selective-reset} process -- 
two selective-reset steps, using a single reset bit;
we thus perform the irreversible part of AC\@.
The dual-selective-reset process is composed of four steps: 
1.-- transfer polarization from \h\ to the far carbon~(\Cone, 
see Figure~\ref{fig:TCE});
2.-- wait for a suitable amount of time, $t_1$, for \h\ to repolarize;
3.-- transfer polarization from \h\ to the adjacent carbon~(\Ctwo);
4.-- wait an additional period of time, $t_2$, for \h\ to repolarize.
Our additional experimental goal here is to bypass Shannon's bound regarding 
conservation of the total
entropy of our 3-spin system, via heat-bath cooling. 

%
%
In the ideal case, 
starting from equilibrium biases, approximately $\{\eps,\eps,4\eps\}$, 
denoted~$\{1,1,4\}$ for \Cone, \Ctwo\ and \h,
respectively, our algorithm produces the following sequence 
of polarizations:
\begin{eqnarray}
\label{Eq:HBC}
\{1,1,4\} \xrightarrow{\rm{SWAP(C1,H)}} \{4,1,1\}
	  \xrightarrow{\rm{WAIT}} 	\{4,1,4\}
\nonumber
\\[3pt]
	  \xrightarrow{\rm{SWAP(C2,H)}} \{4,4,1\}
	  \xrightarrow{\rm{WAIT}} 	\{4,4,4\}
\end{eqnarray}
If, for instance, 
the initial temperature is $300 K$, these final polarization
biases correspond to 
final effective temperatures of 
75K for each carbon and equilibrium (300K) for the proton.
The final information content resulting from 
these final polarizations is  
\begin{equation}\label{I-final}
I_{final}^{approx}
\approx 4^2+4^2+4^2=48 \ ,
\end{equation}
a nearly three-fold
increase, which clearly bypasses the entropy-conservation bound.

\section{Experimental Heat-bath Cooling}

%
%
Our cooling experiments 
were performed 
on standard liquid-state NMR spectrometers
with commercially available 
${}^{13}C_2$-trichloroethylene
(labeled TCE), 
shown in Figure~\ref{fig:TCE},  
in which the hydrogen functions as a reset bit (reset spin) owing to its
relatively rapid relaxation, and the two (labeled) carbons serve
as the computation bits (computation spins).
See Subsection~\ref{subsecMM}
for more details about materials and methods used in our
experiments.

%
%
{}From an algorithmic
point of view, a transfer of polarization can be achieved by
exchanging the states of the two spins, using a SWAP gate as
in~Eq.~(\ref{Eq:HBC}).
The {\em required} transfer of polarization, however, 
is uni-directional as 
we are not concerned with residual polarization (after PT) on the
hydrogen, because it is to be reset, regaining most of 
its initial polarization. 
Therefore, the  
implementation of PT is potentially simpler than 
that of a full (bi-directional) SWAP,
as we explicitly show in Appendix~\ref{app:the-experiment}.
Also note that it is
often inefficient to directly implement PT between
non-adjacent spins due to weak scalar couplings. Step~1 above thus
comprises two sequential steps: 1a.-- PT(H $\rightarrow$ \Ctwo); 
1b.-- PT(\Ctwo
$\rightarrow$ \Cone).
We refer to any pulse sequence implementing nontrivial 
heat-bath cooling (leading to 
cooling of at least two computation spins using one reset spin)
via PT, and WAIT (in our case, 
steps 1a, 1b, 2, 3, and 4)
as POTENT: POlarization Transfer via ENvironment Thermalization.
In our POTENT experiment, 
we implement uni-directional PT using a variant of the 
INEPT (Insensitive Nuclear Enhancement by Polarization Transfer)
pulse sequence~\cite{MF79}. 
Our pulse sequence 
uses exclusively (nonselective) ``hard pulses'', 
namely very short pulses over a wide frequency range, such that all nuclei
of the same species (in our case, the two carbons) 
are similarly affected.

%
%
To estimate the required WAIT times, $t_1$ and $t_2$, one 
must account for the finite ratios of \Tone\ values.
We performed a numerical simulation
of the POTENT pulse sequence 
using a standard relaxation model and \Tone\ relaxation values
which we 
measured in the laboratory. The simulation 
provides an estimated range 
where one should experimentally test for 
the optimal values 
of these two \h\ repolarization delays, 
namely, delays that yield maximal final information content.
For details of the simulation see section~\ref{sec:simulations}.

\subsection{Materials and methods}~\label{subsecMM}
\begin{figure}[h]
 {\centering \resizebox*{0.5\textwidth}{!}{\includegraphics[angle=270]{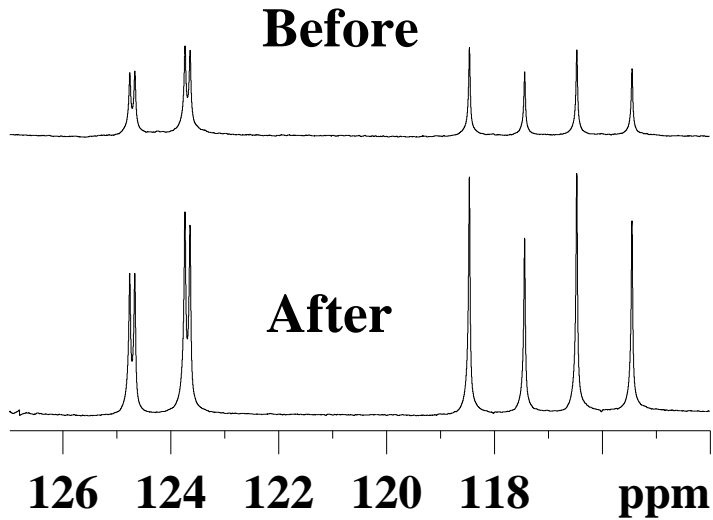}} \par}
 \caption{Carbon spectra before and after 
 POTENT sequence from an experiment performed at Montr{\'e}al in 2002.} 
   \label{fig:montreal}
 \end{figure}

%
%
Our POTENT experiments 
were performed on standard 400-600~\MHz 
liquid-state NMR spectrometers in three different labs,
using different solvents.
The first successful experiment was performed at the Universit{\'e} de
Montr{\'e}al in March~2002 on a Bruker DMX-400
by the authors GB, JMF, TM and YW, together with Raymond Laflamme
from University of Waterloo (Ontario, Canada).
Initial and final carbon 
spectra of the original experiment are shown in Figure~\ref{fig:montreal}.
Only \isotope{C}{13} spectra were recorded, therefore proton polarizations were calculated
according to the simulation model. The calculated information content
was beyond the entropy bound.

%
%
Many POTENT experiments were carried out later on at the Technion and at
Oxford university, with various delays, $t_1$ and $t_2$.
The results presented in the ``Results'' section were obtained
 at the Technion using a Bruker Avance 500 \MHz spectrometer with deuterated
 chloroform (CDCl$_3$) as the solvent.
\isotope{C}{13}$_{2}$-trichloroethylene (TCE) was obtained from CDN Isotopes
(99.2\%\ \isotope{C}{13}) or from Cambridge Isotope Laboratories
 (99\%\ \isotope{C}{13}, diisopropylamine stabilized).
In Montr{\'e}al and in Haifa, the same Bruker pulse sequences 
were used on samples of TCE in deuterated chloroform
(Aldrich, 99.9\%D). At Oxford University, a Varian Inova 600~\MHz spectrometer
was used; 
functionally equivalent pulse programs were applied to TCE samples
in either deuterated chloroform or deuterated acetone. 

%
%
Most experiments were carried out with a paramagnetic relaxation reagent, 
Cr(III)\-acetly\-acetonate (Cr(acac)${}_3$), obtained from Alfa Aesar
 (97.5+\% pure), at a final concentration of about 0.2 mg/mL\@.
The reagent was added to increase the relaxation times ratios, as suggested
 in~\cite{FMW05}; Although Shannon's entropy bound was bypassed in both cases
(with and without reagent), the experimental success was much more robust when
 the reagent was used. The results presented in the ``Results'' section 
were obtained with the relaxation reagent.

%
%
For implementing the PT, high-power, non-selective (``hard'') pulses were
 applied, where the proton and the adjacent carbon, \Ctwo\ 
(see Fig.~\ref{fig:TCE}) were on resonance. 
Selective addressing of one of the two \isotope{C}{13} spins was
achieved by inducing a phase separation using the chemical shift,
while refocusing the scalar coupling evolutions. 
More details can be seen in Appendix~\ref{sec:exp-det}.
Signal averaging (and phase cycling) were not employed, in order to isolate the 
effect of heat-bath cooling.

\section{Results}

\begin{figure}[tb]
\begin{center}
\includegraphics[scale=1]{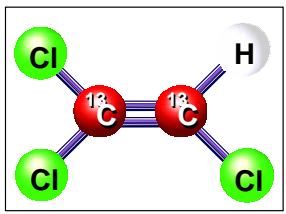}
\end{center}
\caption{
Trichloroethylene labeled with two \isotope{C}{13}. We
denote the leftmost \isotope{C}{13} in this figure as \Cone\ and
the other, neighboring \isotope{H}{1}, as \Ctwo . In our
experiments, the resonance frequencies were 125.773354, 125.772450
and 500.133245~\MHz for \Cone , \Ctwo\ and \h, respectively.
The scalar coupling constants were 201, 103 and 9 Hz
between \Ctwo-\h, \Cone-\Ctwo\ and \Cone-\h, respectively, while
\Tone\ relaxation times were measured at $43 \pm 4\sec$ and
$20 \pm 2\sec$ for \Cone\ and \Ctwo, respectively, and $3.5\pm 0.1\sec$ for \h.
}
\label{fig:TCE}
\end{figure}

\begin{figure*}[h]
\begin{center}
\subfigure[][]{%
  \includegraphics[scale=0.25,angle=270]{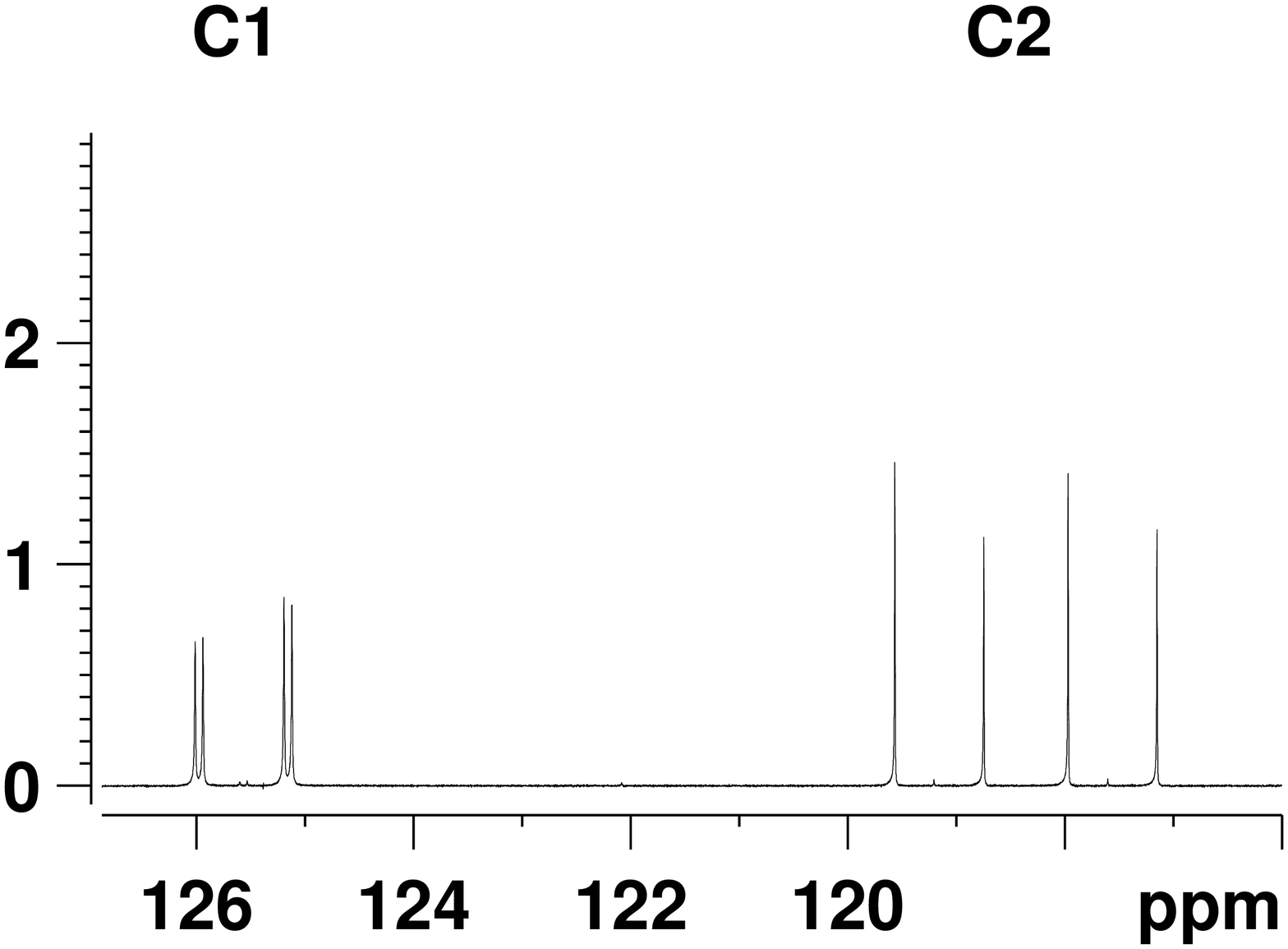}
  \label{fig:zg-C}}
\subfigure[][]{%
  \includegraphics[scale=0.25,angle=270]{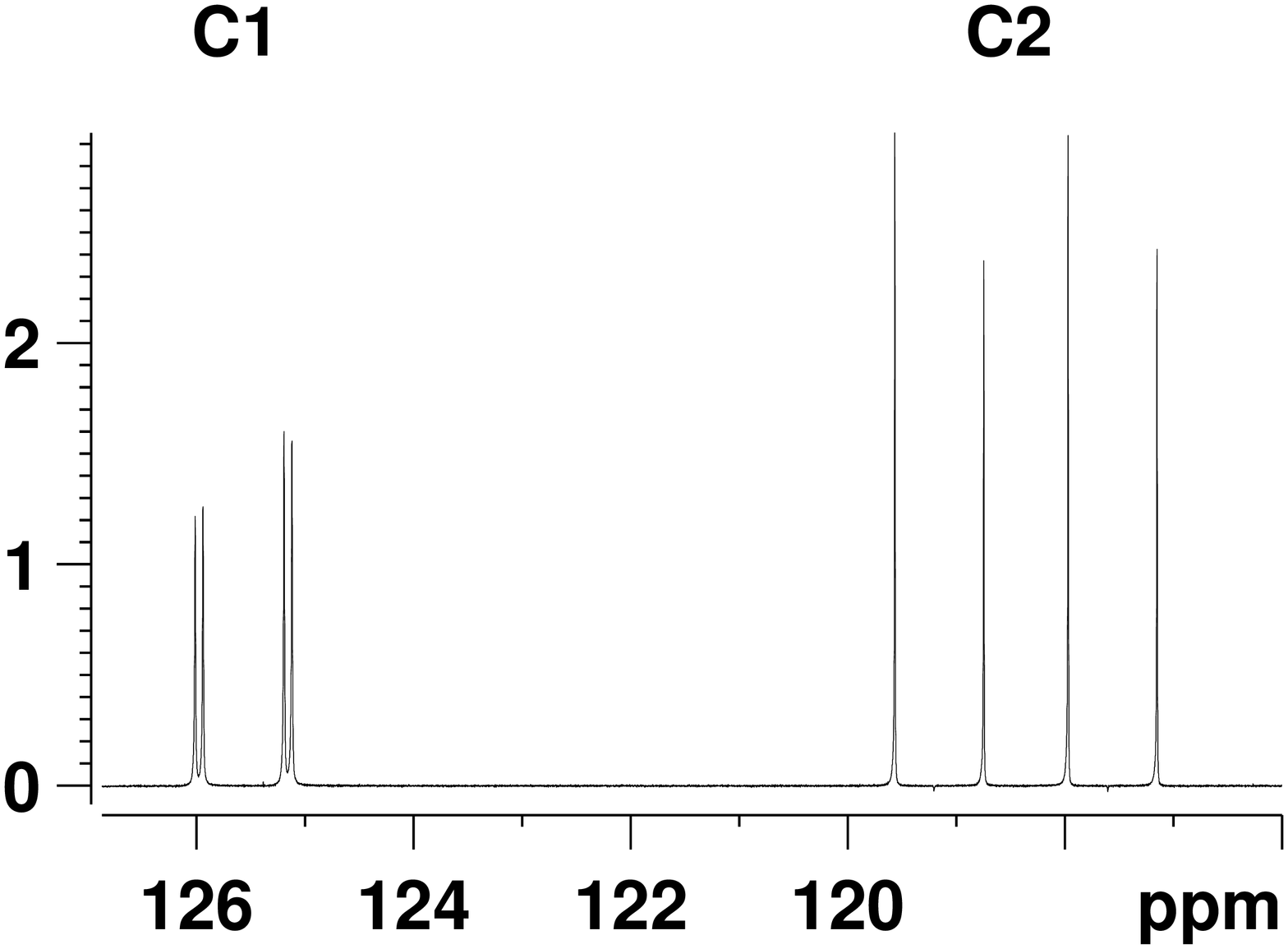}
  \label{fig:after-C}}\\
\subfigure[][]{%
  \includegraphics[scale=0.25,angle=270]{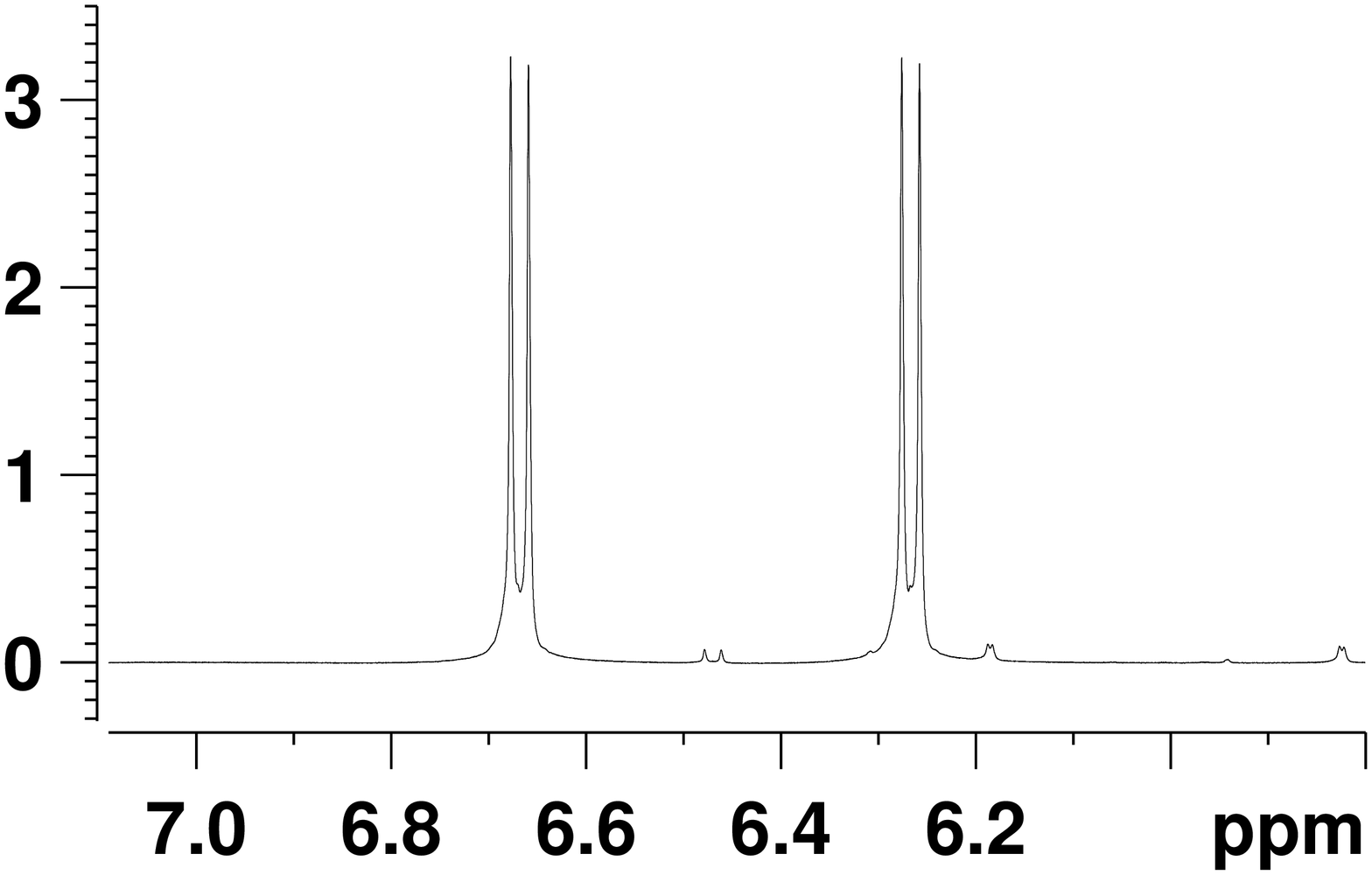}
  \label{fig:zg-H}}
\subfigure[][]{%
  \includegraphics[scale=0.25,angle=270]{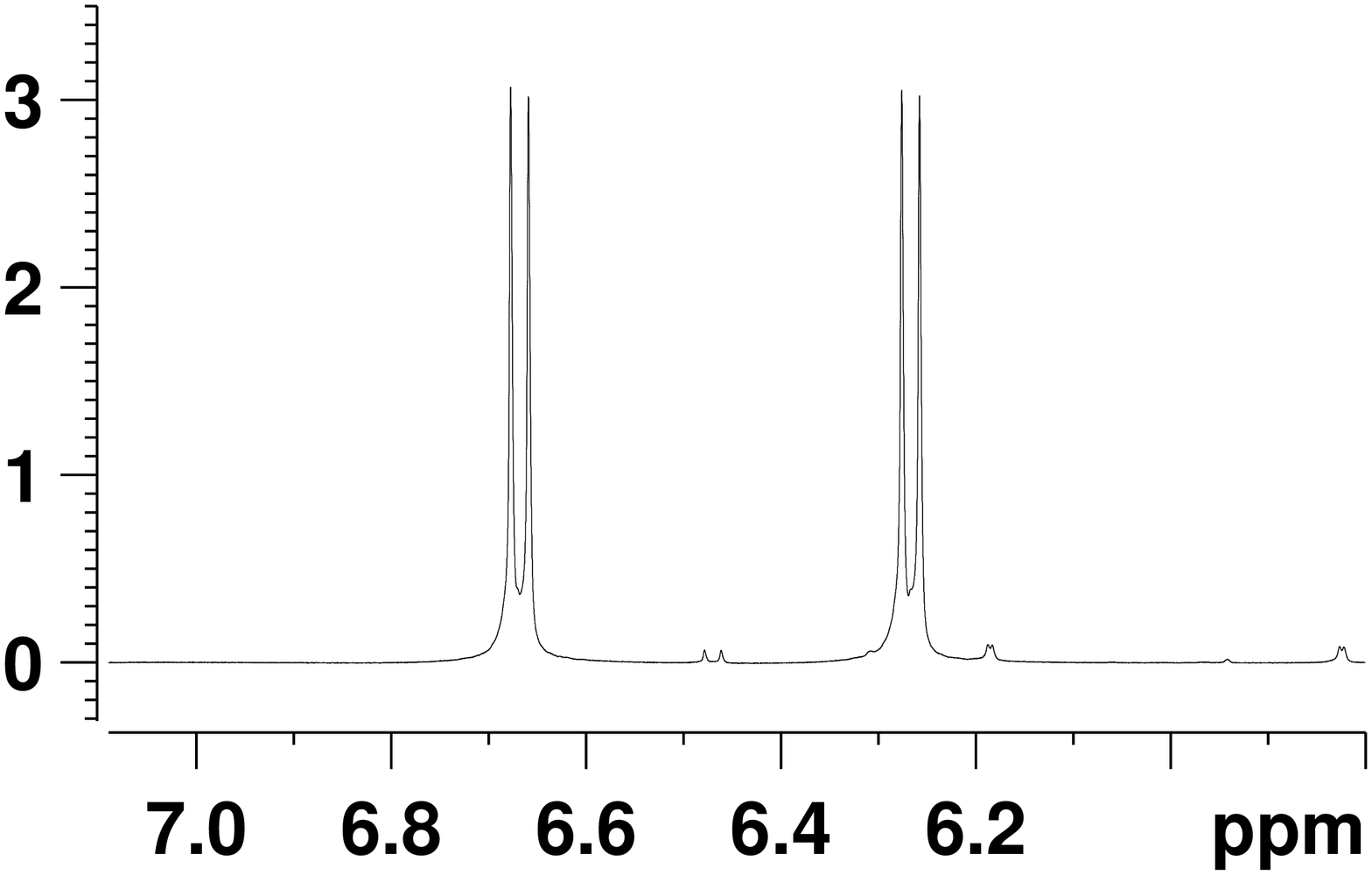}
 \label{fig:after-H}}
\caption{
  Spectra of TCE before and after the heat-bath cooling experiment.  
  Vertical axes denote intensity in arbitrary units, scaled differently 
  for each nucleus.
Figs.\
  (a) and (b) are the \isotope{C}{13}
  spectra before and after the experiment, respectively, with the
  left multiplet being \Cone\ and the right one \Ctwo.  Figs.\
  (c) and (d) are the corresponding \isotope{H}{1} spectra
  before and after the experiment, respectively. The spectrum in
  Fig. (d) was obtained by running the AC experiment
  a second time with the exact same parameters as in
  Fig.~(b), this time observing the \isotope{H}{1} instead of
  the \isotope{C}{13}s by reversing the spectrometer channels.
}
\label{fig:before-after}
\end{center}
\end{figure*}

%
%
We acquired spectra of 
TCE at equilibrium and after the cooling pulse sequence.
Figure~\ref{fig:before-after} displays \isotope{C}{13} and \isotope{H}{1}
NMR spectra for TCE\@. 
Figures \ref{fig:zg-C} and~\ref{fig:zg-H} were
obtained at thermal equilibrium and serve as a reference point for
\isotope{C}{13} and \isotope{H}{1}, respectively.
Figures \ref{fig:after-C} and~\ref{fig:after-H} were
obtained after the cooling pulse sequence.
 For both \isotope{C}{13} nuclei, the increase in polarization bias
 can be intuitively observed by looking at the noticeably higher peaks
 compared with the reference spectrum, while the
 \isotope{H}{1} intensity is only slightly reduced. 
As the polarization bias is directly 
proportional to the area under the relevant peak,
its increase/decrease is more accurately calculated 
via integration.
The experimentally acquired resonance frequencies 
(see the caption of Figure~\ref{fig:TCE})
at room temperature ($296 \pm 1$ K) were used to
 calculate equilibrium biases of $1.000 \pm 0.003$ for
 the carbons and $3.98 \pm 0.01$ for the proton.
The errors are mainly due to the temperature uncertainty,
see Appendix~\ref{app:the-experiment}.
These room-temperature polarization biases 
lead to an initial information
 content of
\begin{equation}\label{I-initial-exp-real}
 I_{initial}^{actual}=17.8 \pm 0.1 \ . 
\end{equation}
A maximal bias identical to the hydrogen initial bias ($3.98$) could be achieved 
for all three spins if the
\Tone\ ratios were infinite and all operations were ideal, leading 
to an ideal final
 information content of
\begin{equation}\label{I-final-exp-ideal}
I_{final}^{ideal}=47.5 \pm 0.1 \ . 
\end{equation}
These values replace the approximate values presented in
Eqs.~(\ref{I-initial}) and (\ref{I-final}), respectively.
Both are calculated using the sum-of-squares of the relevant biases. 

%
%
The spectra after the cooling and the final experimental 
polarization biases (presented in  
Table~\ref{tab:biases}) 
were obtained with the delays
$t_1=8\sec$ and $t_2=12\sec$. 
In the numerical simulation that we performed, 
these delays reside on a plateau, 
in which a wide range of delay combinations
give (more or less) the same highest information content. 
The specific values of
 $t_1=8$ and $t_2=12$ were obtained by experimentally optimizing 
the information content while also minimizing the  
overall duration of the POTENT pulse sequence.

%
%
For each nucleus, the \emph{experimental} final polarization bias 
was obtained by comparing its integrals
before and after  the POTENT pulse sequence. 
The resulting final biases
are $\{1.74 \pm 0.01, 1.86 \pm 0.01, 3.77 \pm 0.01\}$,
corresponding to the final spin-temperatures given in  
Table~\ref{tab:biases}; 
The small experimental error in the biases was obtained by repeating 
the experiment five times under the same conditions.
Calculating the final bias of a specific spin from its integrals 
is equivalent to tracing out~\cite{CJ+02}
the other two spins.
The resulting calculation of the information content, in general,
does not provide the information content, $I$, but a lower bound 
on it\footnote{It is a lower bound due to the subadditivity of 
Shannon entropy.},  
which we denote as $\tilde{I}$. 
Such a calculation provides directly the information content ($I$) 
only when the state of the system constitutes a
\emph{tensor product state} of the three spins,
as in the case above of the initial information content,
and as in the cases of the two simulations described below.

%
%
The lower bound on the final experimental information content is 
\begin{equation} \label{I-final-real}
{\tilde{I}}_{final}^{actual}=20.7 \pm 0.1  \ .
\end{equation} 
This presents an increase of about $16\% \pm 1\%$
over the information content at thermal equilibrium, $17.8$, 
proving that Shannon's entropy bound was indeed bypassed. 

%
%
The ideal final value of $47.5$ is devoid of relaxation constraints, 
and therefore does not provide a reasonable prediction of 
the experimental information content.
To obtain a better prediction of the experimental value we
performed numerical simulations that take into 
account the experimental \Tone\ values
(presented in the caption of Figure~\ref{fig:TCE}). 
Assuming perfect PT and our experimental 
delays  
($t_1=8\sec$ and $t_2=12\sec$), 
the 
simulated information content is 
about 
$ I_{final}^{sim} \approx 29 $.
%
%
The remaining discrepancy between actual and simulated values is still
large; this discrepancy can
largely be ascribed to the low efficiency of the PT steps.
The experimental PT efficiencies were about 
$92\%$ in step 1a, $69\%$ in step 1b, 
and $74\%$ in step 1c  
(see Appendix~\ref{app:the-experiment} for details).
A \emph{practical} simulation\footnote{
Both the ideal simulation and the more practical simulation could be
applied to other molecules. 
}, which goes beyond the ideal simulation by taking 
these imperfect PT efficiencies into account  
 yields  
\begin{equation}\label{I-final-sim-prac}
I_{final}^{prac-sim} \approx 22  
\ , \end{equation}
which is   
much closer to
the obtained experimental result. 
We estimate that the low PT efficiencies are mainly due to  
off-resonance effects, decoherence
(dephasing), and imperfections in the pulse sequence. 
The ideal and practical simulations, 
as well as a few subtle factors that potentially
contribute to the small remaining discrepancy 
between 
$I_{final}^{prac-sim}$ and   
${\tilde{I}}_{final}^{actual}$, are described in section~\ref{sec:simulations}.

%
%
As we see in Table~\ref{tab:biases}, 
both carbons of TCE were cooled considerably, 
well below $200\mathrm{K}$, 
following the application of
POTENT.
Note that, in addition to the experimental final biases, the table also 
includes the ideal and the practical simulated (final) biases.  

%
%
If the goal of the effective-cooling is to reach minimal carbon 
spin-temperatures then 
the final thermalization period, 
namely step 4 of our heat-bath cooling,
should be omitted ($t_2$ set close to 0).
The resulting final temperatures of the 
far carbon~(\Cone) and the adjacent carbon~(\Ctwo)
were 
$145 \pm 2\mathrm{K}$~(\Cone), and $101 \pm 1\mathrm{K}$~(\Ctwo).
For more details on this experiment 
see Appendix~\ref{sec:exp-add}.

\section{Simulations}\label{sec:simulations}
%
%
We simulated heat-bath cooling experiments with Matlab (The MathWorks,
Natick, MA, USA). During the two delays, $t_1$ and $t_2$, each spin 
is assumed to relax according to its experimental \Tone.
If SWAP gates were used for the polarization transfer, then the quantum states
of the two spins involved in a SWAP are simply swapped.
However, when dual-CNOT is used (as explained in Appendix~\ref{sec:exp-det})
 polarization of
the source spin is transfered to the target spin, and the polarization of the
target spin is transfered into classical correlations, so that the resulting 
polarization of the source spin is zero after the transfer.

In the simulation we assume that indeed the resulting polarization of the source
spin is zero after the transfer. However, 
correlation terms between the proton and the carbons were ignored on the
basis of an extended-Markov\footnote{The ``extended-Markov model''
is extremely useful in order to extend far beyond the naive ideal model of
theortical AC --- an infinite relaxation-times ratio (see more details in the post-scriptum).}
 assumption: the proton was considered to some extent as part of the
 memory-less environment, because $t_1$ and $t_2$ are 
sufficiently large compared to 
\Tone(\h). On the one hand, its correlations with the carbons are assumed to
fully cancel out during the reset, but on the other hand, its final polarization
bias after the reset is calculated precisely taking into account the exact reset
time.
Correlation terms among the two carbons (after the first PT from \h\
 to \Cone) cancel out due to the reset of the near-carbon with the help of the
uncorrelated proton.     

\begin{figure}[h]
 {\centering \resizebox*{0.88\textwidth}{!}{\includegraphics{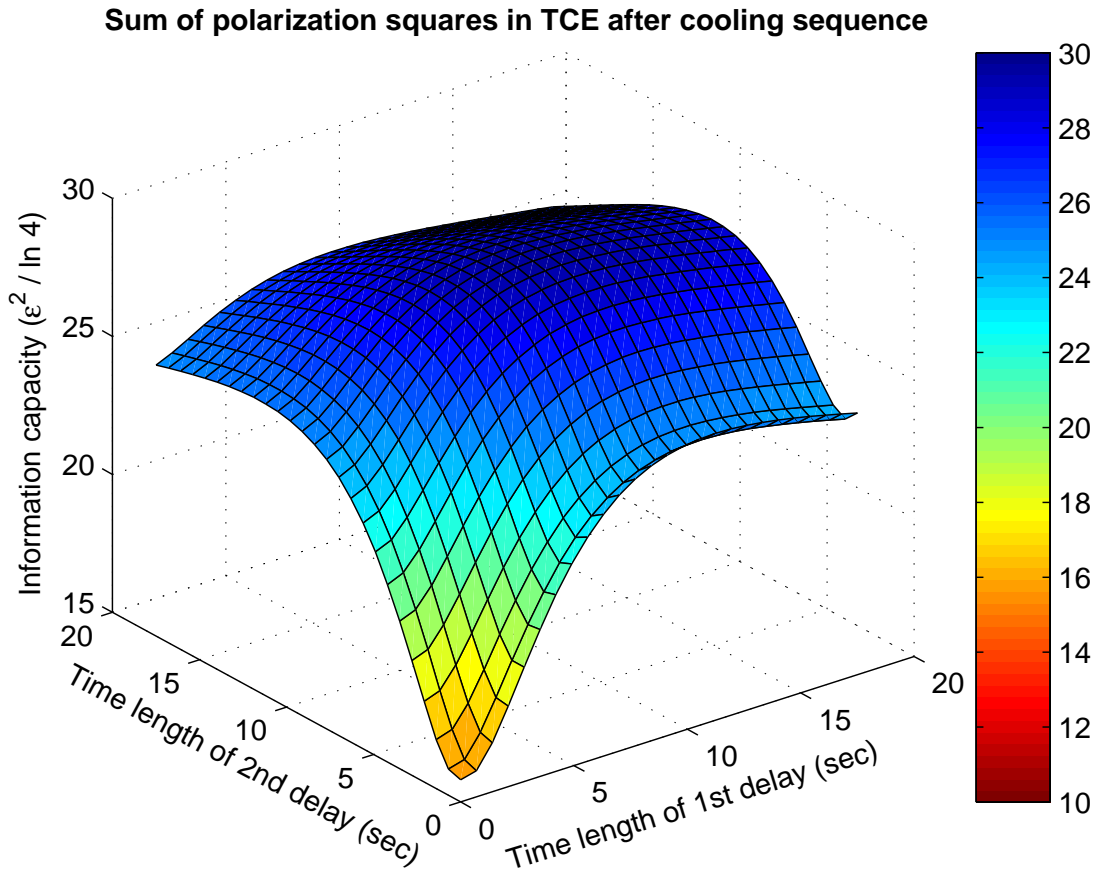}} \par}
 \caption{A simulation of information content (IC), labeled `information 
   capacity', as a function of
   the \h\ repolarization delay times $t_1$ and $t_2$, with the 
   IC represented on the \emph{z}-axis.
   In this simulation we assume perfect polarization transfers.
   The maximum IC value was found numerically at $t_1=9.1\sec$--$10.1\sec$
   and $t_2=8.0\sec$--$8.5\sec$.}
   \label{fig:matlab}
 \end{figure}

The polarization 
biases were calculated as a function of time by the formula
\begin{equation}
\eps^S(t^S)  = (\eps_{init}^S - \eps_{eq}^S )e^{-\frac{t^S}{\Tone^S}} + \eps_{eq}^S,
\label{eq:simulation-eq}
\end{equation}
where for spin $S$, $\eps_{init}^S$ and $\eps_{eq}^S$ are the initial and equilibrium biases,
$t^S$ is the duration, and $\Tone^S$ is the \Tone\ value. The following values were used:
\begin{itemize}
\item For \Cone: $\eps_{init}^\Cone = \eps_{eq}^\h = 3.98 \pm 0.01$, 
     $\eps_{eq}^\Cone = 1.000 \pm 0.003$, $t^\Cone = t_1+t_2$,
     $T_1^\Cone =43\sec \pm 4\sec$
\item For \Ctwo: $\eps_{init}^\Ctwo = \eps_{eq}^\h
	\left(1-e^{-\frac{t1}{T_1^{\h}}}\right)$, 
     $\eps_{eq}^\Ctwo = 1.000 \pm 0.003$, $t^\Ctwo = t_2$,
     $T_1^\Ctwo=20\sec \pm 2\sec$
\item  For \h: $\eps_{init}^\h = 0$, $\eps_{eq}^\h= 3.98 \pm 0.01$, 
	$t^\h = t_2$, $T_1^\h=3.5 \pm 0.1\sec$
\end{itemize}
Biases and corresponding information contents (IC) were obtained for
a fine mesh of delay combinations, in which each delay was varied between $0$
and $5\Tone^\h$ at $2$ \msec\ intervals. The resultant two-dimensional IC
surface for the simplest model (assuming perfect PT steps) is shown in
Figure~\ref{fig:matlab}; Note the broad plateau of near-maximal IC encompassing
a large range of $t_1$ values. The maximal IC of $29.6 \pm 0.7$ was obtained
with $t_1=9.604\sec$ and $t_2=8.239\sec$, which yielded biases of $2.97 \pm
 0.08$, $2.80 \pm 0.08$, and $3.60 \pm 0.03$, for \Cone, \Ctwo, and \h,
 respectively. The errors were derived from Eq.~\ref{eq:simulation-eq} by the
 standard error formula
$$
\Delta \eps = \sqrt{\sum_i\left(\frac{\partial \eps}{\partial x_i}\Delta x_i\right)^2},
$$
where $x_i$ are the relevant variables in each case.

%
%
Imperfect PT was incorporated into the simulation model by adding
our experimental PT efficiencies (see Appendix~\ref{sec:exp-det}), namely $f_1=92\% \pm 2\%$
(initial \h\ to \Ctwo), $f_2=69\% \pm 1\%$ (\Ctwo\ to \Cone), and
 $f_3=74\% \pm 1\%$
(final \h\ to \Ctwo). In Eq.~\ref{eq:simulation-eq}, $\eps_{init}^\Cone$ was multiplied by 
$f_1\cdot f_2$, while $\eps_{init}^\Ctwo$ was multiplied by $f_3$ (The errors 
formulas were adjusted accordingly).
Application of this ``practical'' simulation model to
POTENT yielded an optimal IC with the delays $t_1=11.032\sec$ and $t_2=12.096\sec$;
Note that the second delay is similar to the optimal experimental delay
($t_2=12\sec$), while the first delay is close (within one $\Tone^\h$) to the
optimal experimental value ($t_1=8\sec$). The practical simulation yielded an optimal
IC of $22.3 \pm 0.4$ and corresponding biases of $1.89 \pm 0.06$,
 $1.99 \pm 0.06$, and $3.85 \pm 0.02$ (for \Cone, \Ctwo, and \h, respectively).
 The same IC was obtained by applying the simulation model with
 the optimal experimental delays.

%
%
The IC obtained in our experiment was slightly lower, 
$20.7 \pm 0.1$. 
We
attribute this small discrepancy to the naivete of our simulation model; the 
spins were assumed to relax independently, while we observed significant
cross-relaxation during the extensive delays.

%
%
One may wonder whether the information content achieved
by the two simulations
should be considered as $I$ --- 
the information content, or $\tilde{I}$ --- a lower
bound on the information content. Under the extended Markov assumption, explained
above, the correlations are assumed to cease to exist. Therefore the spins are
in a tensor-product state, and the resulting biases lead directly to the
information content $I$.  If however, one questions the justification of the
extended Markov assumption, and claims that correlations do remain, then the
resulting information content calculated directly from the biases is
$\tilde{I}$, a lower bound on the actual information content (in which part of
the information remains in the correlations).

%
%
The practical simulation model was also applied to the POTENT experiments that
maximize 2-spin ICs, to determine whether bypassing the entropy bound was still
feasible. Of particular interest was the case where only the two computation
spins (carbons) were considered; Such a bypass would be impressive, as both
spins would be significantly enhanced. Perfect PT allows significant
bypasses considering any pair of spins, in particular, the two carbons are 
cooled to {$3.30 \pm 0.06$, $3.82 \pm 0.02$}, with a 2-spin IC of
 $25.5 \pm 0.4,$ about 40\% beyond the equilibrium IC of the spin system.
The practical simulations indicate that the entropy bound can only be bypassed
 in the two (easier) cases where we ignore one of the carbons. In these cases
 the optimal 2-spin information contents are $I(\Cone,\h) \sim 20$,
 and $I(\Ctwo,\h) \sim 19$. We observed both bypasses experimentally
 (see Appendix~\ref{sec:exp-add}).

\section{Discussion}

%
%
We performed experimental effective-cooling of a spin system, implementing an essential step of algorithmic cooling,``heat-bath 
cooling''. Furthermore, this 
is the first experiment designed specifically 
to bypass Shannon's bound on entropy
manipulations, and (as far as we know) 
also the first one to actually bypass~it.
We quantitatively measured the extent 
to which the entropy bound was bypassed, showing a reduction
of the total entropy by about 16\%.
Our dual-selective-reset
experiment combined several 
polarization transfer steps, each between two selected
spins, with 
relatively rapid thermal
relaxation of the ``reset spin''; The
polarization of \textit{two} \isotope{C}{13} nuclei was 
enhanced using
\textit{one} hydrogen in TCE\@.

%
%
Conventional PT techniques such as INEPT (and such as CP and NOE
which are discussed in Appendix~\ref{app:effective-cooling-approaches-in-NMR})
are all aimed at enhancing the signal of less polarized spins by transfer of polarization from nearby more polarized spins. The latter spin (usually 1H) typically relaxes faster than the target spin (commonly 13C or 15N).
Hence, even for 2-spin systems (e.g., labeled chloroform) a
combination of PT and a WAIT step is
sufficient to obtain both a significantly
cooler carbon and a repolarized proton, thus implementing 
a {\em single selective reset} --- the most basic
heat-bath cooling. 

%
%
Let us consider the measurement results after a single PT and WAIT step. 
For example, an application of PT and WAIT to labeled chloroform 
could already bypass 
the entropy bound\footnote{ 
This could also be true for larger symmetric systems which contain several
equivalent protons each bound to a carbon.}.
While we ourselves did not attempt such 
an ``elementary bypass'' using 
\isotope{C}{13}-chloroform,  
we implemented a quite similar 
elementary bypass 
on TCE by 
applying a single PT from the hydrogen to the nearby 
carbon, followed by WAIT\@. 
We bypassed Shannon's entropy bound, first by simulation 
(see section~\ref{sec:simulations}), 
and then also demonstrated it
experimentally (see Appendix~\ref{app:the-experiment})\@.
It is important to note that 
we are not aware of any previously reported elementary
bypass (not even theoretically). Yet, interestingly, the somewhat simpler 
task of bypassing 
S{\o}rensen's unitarity bound was discussed
theoretically~\cite{EP93}.

%
%
Going beyond such elementary bypasses achieved by a single-selective reset, 
the heat-bath cooling implemented here, 
and even more so --- AC, 
achieve less trivial bypasses, since in these algorithms 
{\em pre-selected spins} 
are cooled via {\em multiple} 
selective reset. 
Thus, 
a main value of our work is in the {\em intentional} harnessing 
of selective reset steps for bypassing Shannon's entropy bound as
a new effective-cooling method in NMR spectroscopy.

%
%
One question that might be asked is whether Shannon's 
entropy bound has ever been \emph{unintentionally} bypassed in previous work.
Various PT methods are often combined with rapid 
relaxation (e.g., by adding a paramagnetic salt) 
for improving standard signal
averaging, by reducing the duration of each 
repetition~\cite{MF79,Freeman98}.
At certain moments, this combination may 
increase the polarization 
of the target spin, 
while also partially restoring the
equilibrium polarization of the source spin,
\emph{possibly} bypassing the entropy bound.
However, in standard signal averaging the measurement is not performed
after an intermediate WAIT step, but rather after each PT step. 
Thus, whether anyone unintentionally bypassed 
the entropy bound in any past work 
will probably remain unknown.

%
%
One might ask to what extent the results of the current 
experiment point towards realistic near-future experiments in which 
more significant cooling might be achieved\footnote{This paragraph is rewritten in the postscriptum, taking into account more recent results.}.
Our demonstration  
complements the previously implemented step of AC, 
polarization compression~\cite{CVS01}, and 
\emph{together} with the polarization 
compression, our result  
demonstrates that AC could 
potentially become a practicable means for 
increasing spin polarization in NMR\@. 
Could this potential be fulfilled in the
near future?

%
%
On the one hand, 
there are still several significant obstacles that 
need to be dealt with along the way, 
before significant 
cooling factors 
can be achieved using molecules 
that are relevant to 
some interesting 
applications in NMR spectroscopy. 
First,
sufficiently large \Tone\ ratios are required in order 
to achieve more significant cooling. 
Secondly, 
as the gaps between the ``ideal simulation'' 
and the ``practical simulation'' 
(see Eq.~\ref{I-final-sim-prac})
show, 
improving the PT efficiencies
is also important in order to achieve 
more significant 
cooling. 
Last, implementation of the compression step
is not trivial when the chemical shift between 
the computing spins
is small, as in the case of the TCE molecule.
See some relevant remarks in the Post-Scriptum.

%
%
On the other hand, 
a cooling factor of 1.5 or 2, relative
to the cooling that is achievable today, could already be
significant for various potential applications, e.g., for 
clinical NMR spectroscopy~\cite{RC05}, and such a factor is well 
within the reach of practicable AC; 
unlike TCE, relevant biomolecules, such as amino acids, 
do have large chemical shifts, even at the typically
low fields of clinical NMR spectroscopy (typically up to 3T).
Furthermore, as we explain in the Post-Scriptum, 
high-fidelity gates have recently become available 
for suitable biomolecules containing 2 to 7 
labeled carbons and one or more hydrogens.

%
%
Finally, we would like to briefly repeat
the two connections to 
NMRQC (and to quantum computing in general):
While the algorithms presented 
in~\cite{SV99,BMRVV02,FLMR04,SMW05} 
are classical, they are implemented by quantum gates 
using tools developed in NMRQC, and 
therefore, our results (taken together with the experimental 
polarization compression~\cite{CVS01}) 
hint that spin cooling
for the purposes of magnetic resonance spectroscopy 
might provide the first near-future
application of quantum computation techniques.
In the long run, AC could also become relevant for
generating nearly pure-state spins for NMR quantum computing;
Use of electron spins in the future could lead to a 
breakthrough in experimental AC, as it would provide the high polarization of
 the electron, in addition to its very short reset time
compared to nuclear spins.

\section{Acknowledgements and Post-Scriptum}
\label{sec:acknowledge-and PS}
%
%
We thank Raymond Laflamme  
for helpful discussions and especially for
participating in the initial stages of designing the experiment. 
We thank
 Yael Balasz, Sylvie Bilodeau, Jean-Christian Boileau, Nicolas Boulant,
 Camille Negrevergne and
Tan Pham Viet for useful discussions, suggestions and help in setting
up the experiments.  
We are grateful to Ilana Frank Mor for numerous helpful comments on the manuscript. 
The work of GB and JMF is supported in part by the
Natural Sciences and Engineering Research Council of Canada.
 The work of GB is also supported in part
 by the Canada Research Chair programme and
the Canadian Institute for Advanced Research.
The work of TM, YW and YE is supported by the Israeli Ministry of
Defense. The work of YE, HG, TM and YW is supported by
the Institute for Future Defense Research at the Technion. JAJ and
LX thank the UK Engineering and Physical Sciences Research
Council and Biotechnology and Biological Sciences Research Council for
financial support.

%
%
When we first submitted this paper to Science and then to PNAS 
in 2005 (See also~\cite{POTENT}), the path
to near-future applications seemed quite long,
as we explain above.  
In hindsight, these results heralded further
 experiments, where more significant cooling was achieved.
Our demonstration complemented the previously implemented step of AC,
polarization compression~\cite{CVS01}, and was recently repeated with
biomelecules~\cite{EGMW11}. The combination of both steps, initially in solid-state
NMR, and recently also in liquid-state NMR~\cite{Atia-MSc-Thesis,AEMW-2}, demonstrates that AC could
potentially become a practicable means for increasing spin polarization in NMR. Could this potential be fulfilled in the near future?

We wrote, in the first public version of this current paper~\cite{POTENT}, 
that the irreversible step of AC is first
done in this work (which was correct then). However, 
shortly after this manuscript was
submitted, a closely related work
was described in~\cite{BMR+05} in which multiple 
selective reset steps were 
also done (independently of our work); 
In that work, AC was performed by combining multiple selective-reset 
 steps with a compression step on a specially designed spin-system 
containing 
one active reset spin and three computer spins, 
using \emph{solid-state} NMR\@.
Their system contained
\isotope{C}{13} labelled malonic acid in a single crystal containing mostly 
unlabelled malonic acid molecules.
In that work, however, Shannon's bound was not bypassed.
Subsequent work by the same 
group~\cite{RMBL08}, published three years 
after we originally submitted the first revision of this paper
describing our success at bypassing Shannon's bound,  also bypassed the entropy bound. This was done using the same solid-state system they had
previously used. 

As expected, we recently achieved trivial bypasses with \isotope{C}{13}-chloroform at 500~\MHz, e.g. the total entropy of this 2-spin system was reduced by about
 20\%, following PT from the proton to the carbon and a long delay of 
 $\sim 7$\Tone(H), which allowed the proton to regain most of its equilibrium polarization (while the carbon retained a cooling factor of about~2).

%
%
Recent experimental 
``optimal control'' methods 
in NMR~\cite{KR+05,TV+09} 
are useful for generating high-fidelity gates. 
This progress, and
the recent experimental heat-bath cooling of 
labeled amino acids~\cite{EGMW11},   
combined with 
recent theoretical algorithmic cooling results~\cite{SMW07,EMW11},
suggest that cooling \isotope{C}{13} spins in
bio-molecules  
by a factor of $\approx2$ relative to the cooling which is
achievable today, is likely 
within a few years. In particular, the compression step
was already successfully implemented using optimal control tools both in 
solid-state NMR~\cite{RMBL08} and in liquid-state NMR\footnote{
Experimental multi-cycle AC of labeled TCE, using optimal control, 
succeeded in bypassing the entropy bound
on a single carbon, cooling it beyond $\sqrt{18}$. See more details in
Appendix~\ref{app:post}, among additional Post-Scriptum details}.
In recent years AC was found useful in various directions~\cite{RMM07,HRM07,
WHRSM08,BP08,LPS10,DRRV11,BPM+11,SBR+11,CML11,Renner12,Blank13,BCC+07,Lloyd14,XYX+14} 
and some of the tools and results presented here
could be relevant elsewhere as well.

\newpage

\printbibliography

\newpage

\newpage

 \begin{figure}[h]
 {\centering \includegraphics[scale=1]{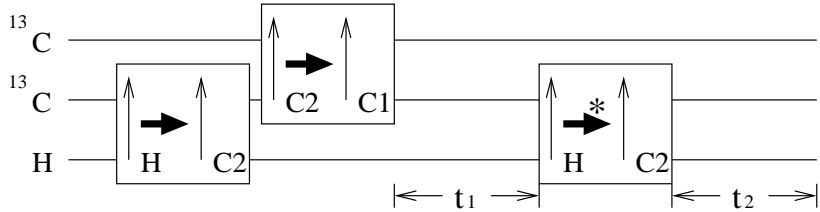} \par}
 \caption{A block diagram of the complete experiment, with \Cone\
   represented at the top line.  The arrow boxes denote polarization
   transfers in the direction of the arrow. The periods $t_1$ and
   $t_2$ are the variable delay times in which we wait for \h\ to repolarize.}
 \label{fig:full-experiment}
 \end{figure}

 \begin{figure}[h]

 {\centering \includegraphics[scale=1]{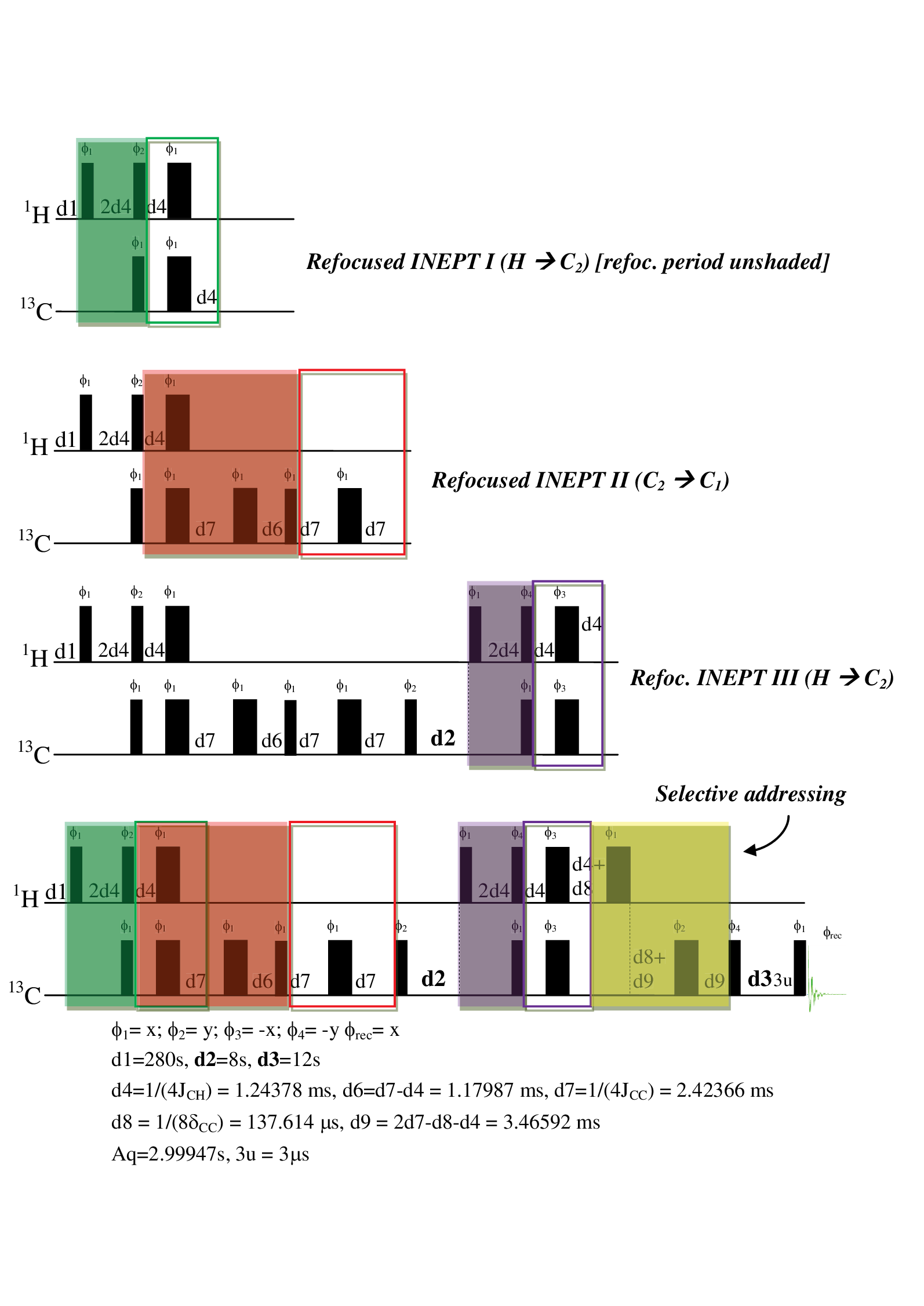} \par}
 \caption{Annotation of the POTENT pulse sequence:~(a),~(b) and~(c) highlight 
 the first, second and third refocused-INEPTs; the first half is shaded, and the
  refocusing period is outlined by a rectangular border of the same color. 
In~(d), the complete POTENT, cf. Fig.~\ref{fig:pulse-sequence}; the yellow block marks selective inversion of \Cone.}
 \label{fig:pulseq-annot}
 \end{figure}

 \begin{figure}[h]
 {\centering \includegraphics[angle=270,scale=0.56]{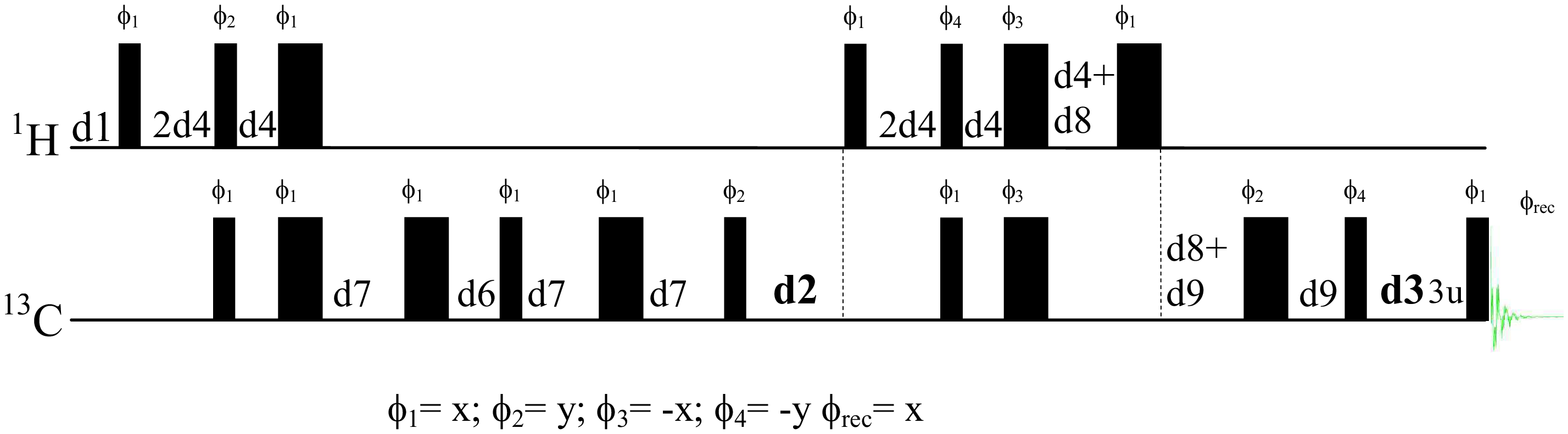} \par}
 \caption{The complete POTENT pulse sequence. 
Narrow bars and rectangles represent $90^\circ$ and $180^\circ$ rotations.
See caption of Fig.~\ref{fig:pulseq-annot} for values of the various
delays. The \h repolarization delays d2 and d3 are denoted in the text by
 $t_1$ and $t_2.$}
 \label{fig:pulse-sequence}
 \end{figure} 
 
 \def\figscale {0.3}
 \begin{figure}[h!]
 {\centering
 \begin{tabular}{|c|c|c|}
 \cline{2-2} \cline{3-3} 
 \multicolumn{1}{c|}{}&
 Step&
 Spectrum after the step\\
 \cline{2-2} \cline{3-3} 
 \hline 
 1&
 \resizebox*{\figscale\textwidth}{!}{\includegraphics{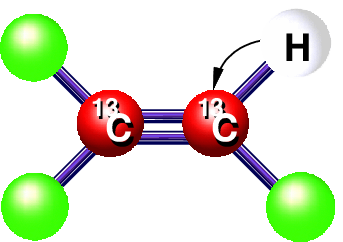}} &
 \resizebox*{\figscale\textwidth}{!}{\includegraphics{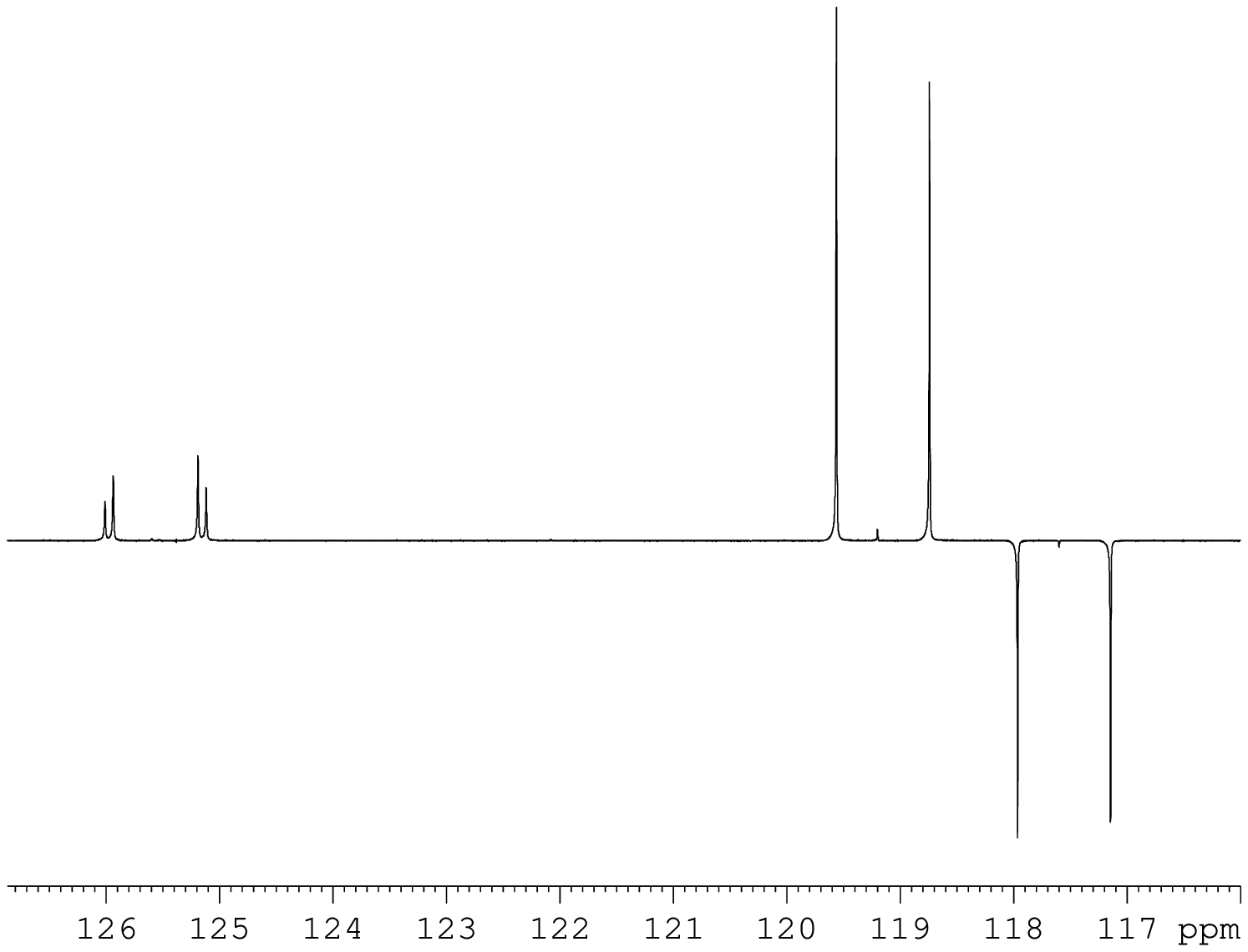}} \\
 \hline 
 2&
 \resizebox*{\figscale\textwidth}{!}{\includegraphics{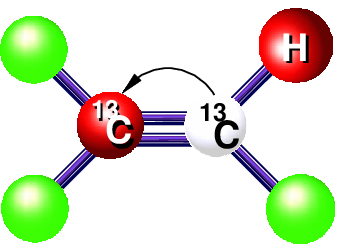}} &
 \resizebox*{\figscale\textwidth}{!}{\includegraphics{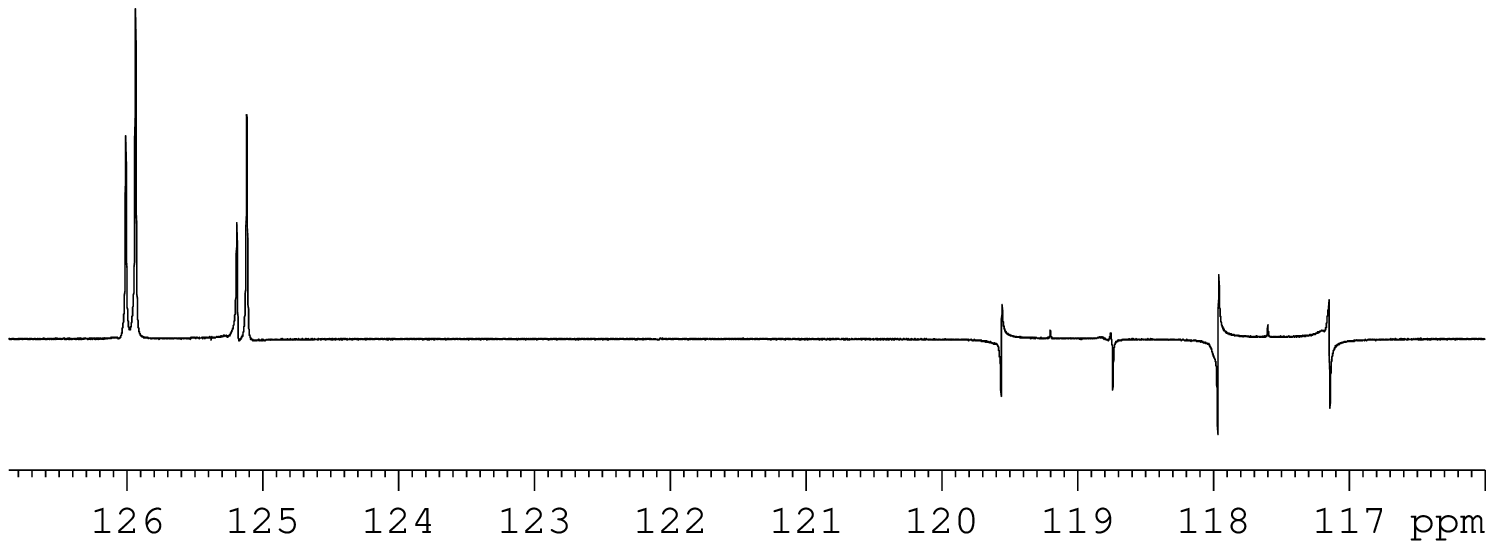}} \\
 \hline
 3&
 \resizebox*{\figscale\textwidth}{!}{\includegraphics{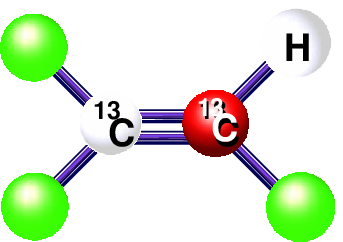}} &
 \resizebox*{\figscale\textwidth}{!}{\includegraphics{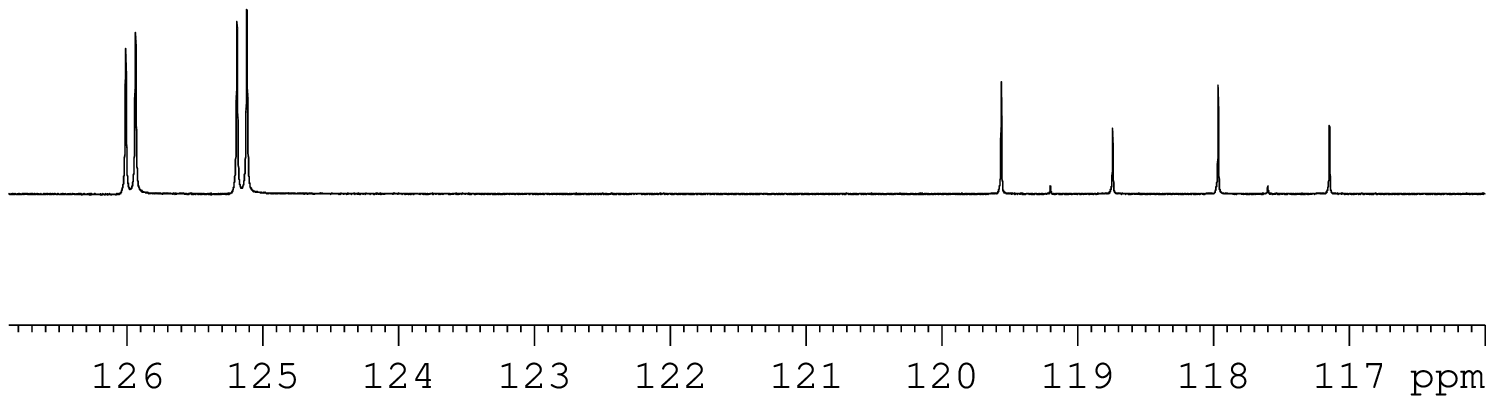}} \\
 \hline 
 4&
 \resizebox*{\figscale\textwidth}{!}{\includegraphics{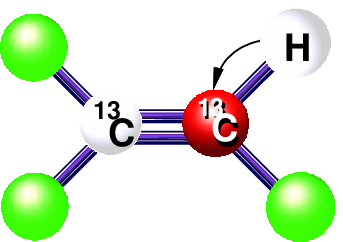}} &
 \resizebox*{\figscale\textwidth}{!}{\includegraphics{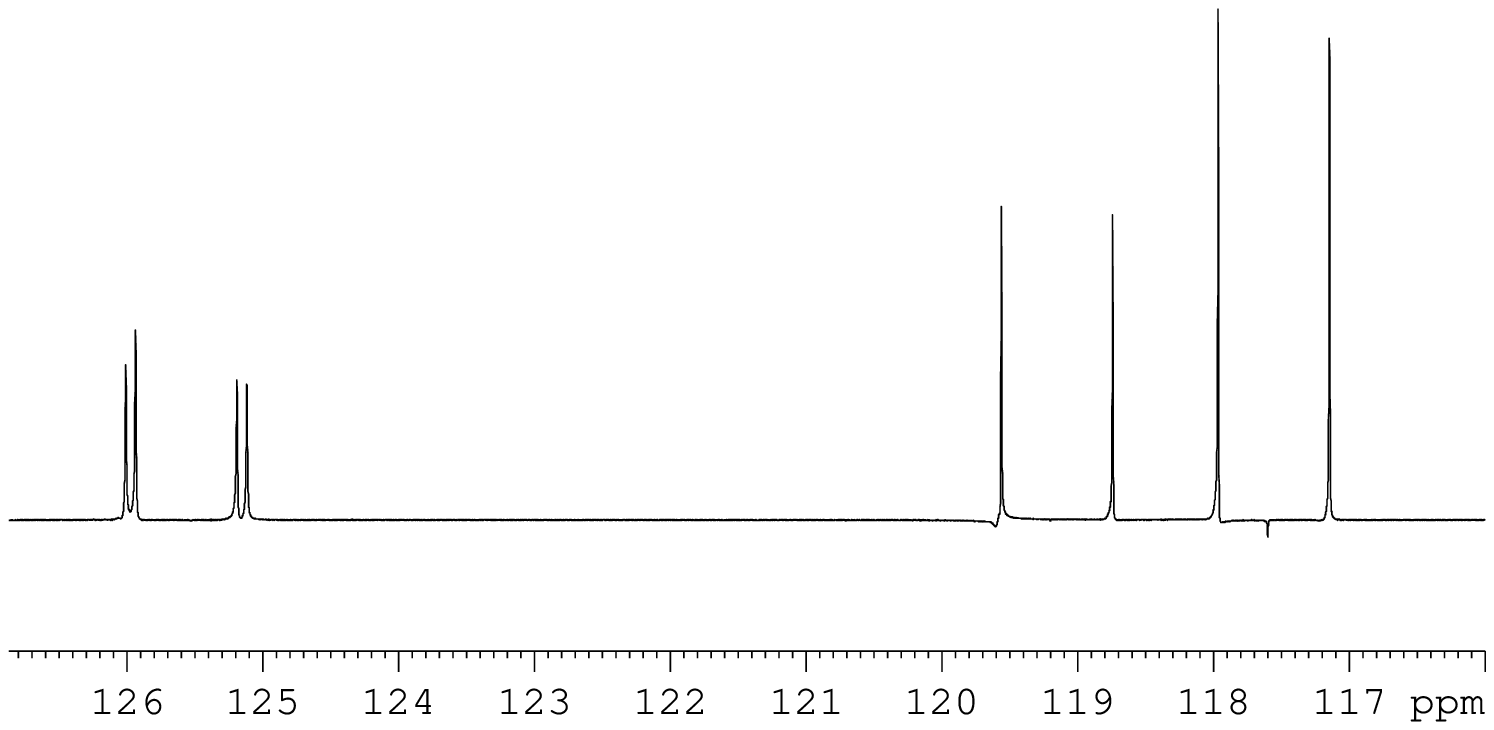}} \\
 \hline
 5&
 \resizebox*{\figscale\textwidth}{!}{\includegraphics{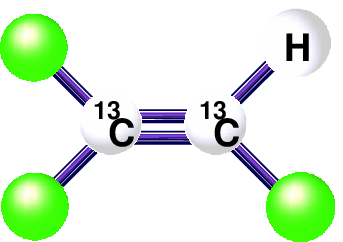}} &
 \resizebox*{\figscale\textwidth}{!}{\includegraphics{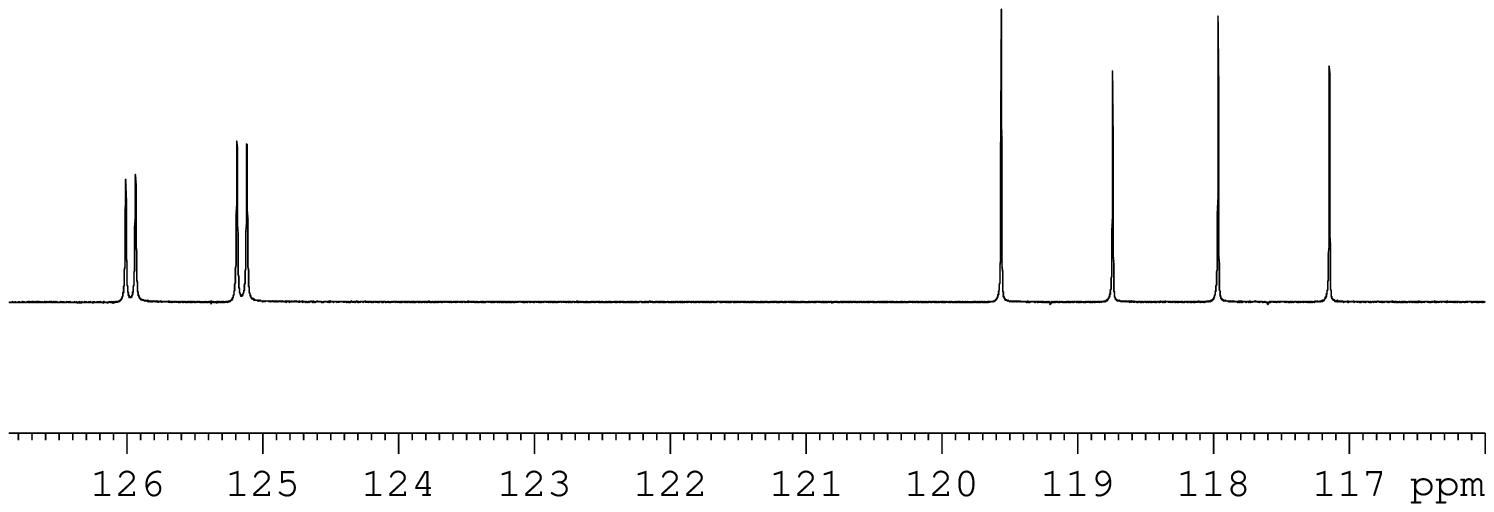}} \\
 \hline
 \end{tabular}\par}
 \caption{The steps of the cooling experiment and the resulting
   \isotope{C}{13} spectra after each step.
 }
 \label{fig:spectra}
 \end{figure}

\begin{singlespace}
\begin{table*}[ht]
\caption{Initial and final polarizations, and final spin temperatures of each 
 bit in TCE for the AC experiment shown in Figure~\ref{fig:before-after}.}
\begin{tabular*}{\hsize}{@{\extracolsep{\fill}}lllclc}
\multicolumn1c{spin}
&\multicolumn1c{Initial bias~($\eps$)}
&\multicolumn1c{Sim bias}
&\multicolumn1c{Prac-sim bias}
&\multicolumn1c{Final bias}
&\multicolumn1c{Final spin T~(K)}\\
\hline\hline
\Cone\ (far) & $1.000 \pm 0.003$ & $2.87 \pm 0.08$ & $1.96 \pm 0.06$ & $1.74 \pm 0.01$ & $170\pm 1$
\\
\Ctwo\ (adjacent) &
$1.000 \pm 0.003$ & $2.41 \pm 0.09$ & $1.90 \pm 0.06$ & $1.86 \pm 0.01$ & $159\pm 1$
\\
\h &
$3.98\pm 0.01$  & $3.85 \pm 0.02$ & $3.85 \pm 0.02$ & $3.77 \pm 0.01$ & $312\pm 1$ \\
\hline
\end{tabular*}
The resonance frequency of each nucleus was used to compute
its bias at thermal equilibrium at the room temperature of
$296\pm 1$ K in which the experiment was run.
Precise values for the final polarization of each nucleus were
obtained by comparing the integrals of the peaks,
before and after the POTENT process; 
This method of 
calculation provides the final bias of each spin by tracing-out
the other two spins (see~\cite{CJ+02}).
The simulated
values, in the third and fourth columns,
were computed by taking into consideration empirical \Tone\ relaxation
times. 
The results in the third column assume perfect 
polarization transfers,
while the results in the fourth column use empirical
transfer efficiencies. 

\label{tab:biases}
\end{table*}
\end{singlespace}

\appendix 

\section{Effective-cooling approaches in NMR}\label{app:effective-cooling-approaches-in-NMR}
%
%
When a spin-half particle is placed in a constant magnetic field and
coupled to a thermal bath, the probabilities of finding it in the
two spin states $\ket{\pm\frac{1}{2}}$ are given by the Boltzmann
formula
$$P_{\pm\frac{1}{2}}=\frac{1 \pm \eps}{2}=
\frac{\exp(-E_{\pm\frac{1}{2}}/k_BT)}{Z},$$ where
$E_{\pm\frac{1}{2}}$ are the energies of the two states,
$k_B$ the Boltzmann constant, $T$ the bath temperature, $Z$
a normalization factor (the partition function), and 
$\eps \equiv P_{\frac{1}{2}} - P_{-\frac{1}{2}}$ is the population
bias or polarization bias.
The equilibrium population bias is therefore $\eps=\tanh(\Delta
E/2k_BT)$, where $\Delta E$ is the energy gap between the two
states. At high temperatures this simplifies to $\eps
\xrightarrow{k_BT \gg \Delta E} \Delta E/2k_BT$.
The equilibrium population bias in liquid-state NMR systems at 
room temperature is typically rather
small --- below $10^{-4}$. 

%
%
As a result of the small polarization bias, the obtained signal in
dilute solutions at room temperature is often 
small if a single measurement 
(a single scan) is performed. 
Signal averaging over many scans 
is then required to attain sufficient signal-to-noise ratio.
To reduce the experimental duration, a wide range of effective-cooling
techniques were developed to increase population biases, without 
cooling the system. Some
of these methods are briefly discussed below. The effects can be described 
in terms of the final bias achieved, or equivalently in terms of a
spin temperature, defined by
$$T\triangleq\frac{\Delta E}{2k_B\tanh^{-1}\eps} \xrightarrow{\eps\ll1}
\frac{\Delta E}{2k_B\eps}.$$

\paragraph{Conventional polarization transfer.}

Elementary polarization transfer (PT) techniques such as INEPT are widely
used in NMR studies~\cite{Sorensen89}, leading to modest
polarization enhancements for the spins of interest.
These techniques usually rely on
transferring polarization from a hydrogen nucleus, with a relatively
large bias, to a directly bonded carbon or nitrogen nucleus with a
much smaller bias.  These techniques are simple to implement, but
cannot give large enhancements.

There are other techniques such as cross-polarization (CP),
continuous CP, and NOE that can be used similarly to INEPT for transferring
polarization from protons to carbons and other spins.
In contrast with INEPT, where the PT is through bond, NOE is manifested
through space, and is therefore simple to implement yet less efficient (limited
 to 75\% of the proton polarization).

\paragraph{Dynamic Nuclear Polarization (DNP).}
 Dynamic Nuclear Polarization~\cite{Overhauser53}, or DNP, refers to 
processes that transfer polarization from the spins of unpaired electrons to
 nearby nuclear spins, e.g. by saturating electronic transitions with microwave irradiation. DNP can yield polarizations close to 1 for solid samples at very
 low temperatures after prolonged
 irradiation. Much of the hyperpolarization is preserved when rapidly
 dissolving the sample~\cite{AFGH+03}, however dissolution DNP is slow and
 requires complex equipment, and therefore it cannot be used to refresh spins.

\paragraph{Electron Nuclear DOuble-Resonance (ENDOR).}
While not a spin-cooling approach \emph{per se}, electron nuclear
 double-resonance (ENDOR)~\cite{Slichter90} takes advantage of the high
 polarization of electron spins to resolve NMR lines by detecting the resonances
 of nearby electron spins, e.g. in paramagnetic centers. The signal of a
 particular electron transition is first removed by saturating it with microwave
 radiation and then restored by irradiating at a radiofrequency around the
 hyper-fine coupling to exchange the population of electron and nuclear energy
 levels. The matching of electronic and nuclear transitions requires complex
 equipment and precise temperature control. 

\paragraph{Hyperpolarized gases.}
Optical pumping approaches achieve extremely high spin polarizations in noble
 gases, most notably xenon~\cite{GUTP05}.
A small portion of the hyperpolarization can be transferred to molecules of
 interest by co-condensing them with the noble gas and using simultaneous
 irradiation~\cite{NSR+96,FSH98,VLSVC01}. By this process, carbon and proton
 spins in chloroform were equally enhanced~\cite{VLSVC01}, and larger
 enhancements of about 20--40-fold were obtained at low temperature and field
 (200K and 1.4T)~\cite{FSH98}. The transfer of polarization from xenon to
 \isotope{H}{1} or \isotope{C}{13} is non-selective and sometimes also
 non-uniform~\cite{FSH98}. The process is complicated, requiring a laser array,
 an additional low magnetic field (0.1T), and an optical pumping chamber. 

\paragraph{Parahydrogen induced polarization (PHIP).}
At low temperature (below about 85\,K) in the presence of a suitable catalyst,
hydrogen gas assumes its rotational ground state, in which the two hydrogen
nuclei are in a spin singlet state, known as \textit{para}-hydrogen.
This state is preserved upon warming to room temperature, and can be added
across a double or triple bond~\cite{AJB+04}; the parahydrogen
induced polarization (PHIP) is very high, sometimes approaching 
unity~\cite{AJB+04}. PHIP requires special equipment and cannot be generated in
 vivo.

\paragraph{Polarization compression and algorithmic cooling.}

In this appendix we focus so far on techniques that cool to the temperature of 
the high-bias spin. 
In the main text we describe going below the temperature of
the high-bias spin using polarization compression and algorithmic cooling.

\paragraph{Beyond simple signal averaging.}
Last but not least, signal averaging may be enhanced by PT. First, when the
source-spin for the PT is as polarized as the target spin (or even if it 
is less polarized than the target spin) by employing, 
as in algorithmic cooling,  
the idea of fast reset: Assuming the source and the target spins have the same
polarization, but the source spin resets ${\cal R}$ times faster, PT is then
simply used for obtaining a rapid recovery after each scan. Comparing this idea 
to algorithmic cooling was not done yet and is left for future
research. 

A similar case, in which the source spin is
also useful due to its higher polarization, 
was demonstrated experimentally~\cite{MF79} and discussed in~\cite{Derome87}.
Due to the high relevance of this case to this paper,
we discuss it in the ``Discussion'' section of the paper.

An important aspect of AC and its special case heat-bath cooling
is the possibility of synergy with many other
methods of signal enhancement: signal
averaging, physical cooling, increasing the magnetic field, etc.

\section{Reversible compression and algorithmic cooling}~\label{app:reversible-compression-and-AC}

\subsection{From loss-less in-place data compression to polarization
compression}~\label{app:loss-less-compr-to-polarization-compr}

%
%
Schulman and Vazirani (SV)~\cite{SV99} were the first to suggest that 
loss-less in-place data compression can lead to polarization compression.
Consider a bit-string of length $n$, such that the probability distribution
is known and far enough from the uniform distribution. 
One can use data compression to generate a shorter string, 
say of $m$ bits, such that the entropy of each bit is much closer to one.
As a simple example~\cite{Mor07}, 
consider a four-bit-string which is distributed as follows:
$p_{0001}=p_{0010}=p_{0100}=p_{1000} = 1/4$,
with $p_i$ the probability of the string value $i$ 
(the probability of any other value is zero).
The bit-string can be compressed, via an algorithm, into a 2-bit string that
holds the binary description of the location of ``1'' in the above four strings.
 
A similar loss-less, in-place, data compression process could generate an
 output of the same length $n$ as the input, such that the entropy is compressed
into the last two bits. For instance, logic gates that operate on the bits can perform the permutation, 
$\{0001 \rightarrow 0000; 0010 \rightarrow 0001; 0100 \rightarrow 0010;    
1000 \rightarrow 0011\}$,      
while the other input strings (whose probability is anyhow zero) 
transform to output strings in 
which the two most significant bits are not zero; for instance
$1100 \rightarrow 1010$. 
The entropy is now fully concentrated on
the two least significant bits, thus the above process implements 
a simple case of data compression. The two
 most significant bits have zero entropy; if 
these two bits were spins, the process would have made them extremely cold. 

%
%
In order to gain some intuition about the design of logic gates
that perform such entropy manipulations when using nuclear spins, 
consider a closely related scenario
 (first considered by von Neumann):
fair coin flips can be extracted, given a biased coin, by taking a pair of
biased coin flips, with results $a$ and $b$, and using the value of $a$
\underline{conditioned on} $a \ne b$. A simple calculation shows that $a=0$
and $a=1$ are now obtained with equal probabilities, and therefore the entropy
of coin-flip $a$ is increased in this case to 1; as we soon shall see,
this means that its temperature is increased (to infinity).
The opposite case, the probability distribution of $a$ given that 
$a=b$, results in a highly determined coin flip; namely, 
a (conditioned) coin-flip with a higher bias or lower entropy. 
A gate that flips the value of $b$ if (and only if) $a=1$ is called a Controlled
-NOT gate (CNOT). Consider the value of $b$ following the CNOT, namely $b_f$;
$b_f=1$ implies that $a \ne b$ prior to the gate; the final entropy of $a$ is
 then 1. On the other hand, $b_f=0$ implies that $a = b$ prior to the gate;
 the final entropy of $a$ in this case is lower than its initial value. 
The similar scenario of flipping two identical coins is more relevant to 
spin-entropy manipulations discussed below.

%
%
For quantum two-level systems (e.g., spin-half nuclei) there is a simple
 connection between temperature, entropy, and population probability. The
 process of increasing the polarization bias (reducing the entropy) without
 cooling the thermal-bath is known as ``effective-cooling'' (Appendix~A). 
We can conclude that the two most significant bits in the first example 
got much colder (actually, to ZERO temperature) during data compression.
The second example is directly relevant to nuclear spins, and we see that this
 process can cool some of them. The CNOT gate can be combined with other simple
 gates in a useful cooling subroutine.

%
%
SV~\cite{SV99} identified the importance of the low-entropy bits resulting from
in-place loss-less data compression. Physical spin-half nuclei may be similarly
 cooled by data compression algorithms. Consider a linear molecule with $3$
 adjacent spin-half nuclei; At thermal equilibrium at room temperature and a
constant magnetic field, the bits on each molecule are essentially uncorrelated.
Furthermore, in the liquid state one can also neglect the interaction between
 molecules. It is convenient to write the probability distribution of a single
spin at thermal equilibrium using the ``density matrix'' notation
\begin{equation} \label{init-1b-state}
  \rho_{\eps} = 
  \begin{pmatrix}
               p_{\uparrow}  & 0 \\
               0                & p_{\downarrow}
  \end{pmatrix}  =
  \begin{pmatrix}
               (1+\eps)/2 & 0 \\
               0                & (1-\eps)/2
  \end{pmatrix} 
\ , \end{equation}
since these two-level systems are of a quantum nature (namely, these are quantum
 bits --- qubits), their states are not limited to a classical probability
 distribution over `0' and `1'. 
We consider now the classical case where $\rho$ contains only diagonal elements
 that describe a conventional probability distribution.
At thermal equilibrium, the state of $n=2$ uncorrelated qubits with the same
polarization bias is described by the density matrix 
$\rho_{\mathrm{init}}^{\{n=2\}} = \rho_{\eps} \otimes \rho_{\eps}$,
where $\otimes$ means tensor product.
The probability of the state `00', for instance, is $(1+\eps)/2 \times 
(1+\eps)/2  = (1+\eps)^2/4$ (etc.)
Similarly, the initial state of a $3$-qubit system of this type, at thermal
 equilibrium, is  
$\rho_{\mathrm{init}}^{\{n=3\}} = 
\rho_{\eps} \otimes \rho_{\eps} \otimes
\rho_{\eps}$.
This state represents a thermal probability distribution, such that the
probability of the classical state `000' is $p_{000}=(1+\eps_0)^3 / 2^3$, etc.
In reality, spins generally have different initial biases, but as long as these
 differences are small (e.g., for homonuclear spins), we ignore them.
Sufficient chemical shifts (resonance difference) for spins of each type are assumed,
to allow selective addressing.

%
%
SV analyzed the cooling of such systems using various tools of data compression.
Their technique and similar ones are now called ``reversible polarization compression''.
Some ideas of SV were already explored a few years earlier by 
S{\o}rensen~\cite{Sorensen89}, who analyzed effective-cooling of spins,
with no connection to data compression. The entropy bound (due to Shannon) tells
 us that the total entropy of the system cannot be reduced by such processes.  
A tighter bound~\cite{Sorensen89} applies when only unitary operations are allowed. 

%
%
\underline{\bf Example:} 
The polarization bias of a single spin (bit)
in an $n$-spin molecule at room temperature, 
assuming all polarization biases are initially $\eps$
and are much smaller than 1, cannot be reduced~\cite{FLMR04} 
by more than a multiplicative factor of $\sqrt{n}$:
The total entropy of such a molecule,   
$H(n) = n(1-\eps^2/\ln 4)+O(\eps^4)$,
is compressed so that $n-1$ spins have maximal entropy; the remaining
spin satisfies 
at best  $H({\rm single}) = 1-(\sqrt{n}\eps)^2/\ln4 + O(\eps^4)$,
so that 
\begin{equation}\label{reversible-cool-limit}
\eps_{final} \approx \sqrt{n}\eps
\end{equation}
 (the approximation is valid as long as $\eps_{final} \ll 1$).
Spins actually cannot be cooled to the extent allowed by Shannon's bound 
due to S{\o}rensen's unitarity bound~\cite{Sorensen89}, 
since arbitrary entropy-preserving manipulations are not
physically possible.

\subsection{Algorithmic Cooling --- a brief review} ~\label{app:AC-brief-review}

Boykin, Mor, Roychowdhury, Vatan and Vrijen suggested in 2002  
a effective-cooling technique, which they named  
\emph{Algorithmic Cooling (AC)}~\cite{BMRVV02}, or more specifically,
heat-bath AC\@.
Controlled interactions with a heat bath allow, theoretically, cooling much
 beyond entropy preserving processes. In order to pump entropy out of the
 system, AC employs designated computation spins together with rapidly relaxing
 reset spins, repeating the following steps several times: entropy compression, entropy shift onto reset spins using PT, and entropy removal from the system
(reset spin repolarization).

%
%
The concept of AC led to practicable AC (PAC)~\cite{FLMR04} for cooling
 \emph{small molecules}. PAC schemes use PT steps, reset steps, and 
a basic 3-spin compression step termed \emph{3-bit-compression
(3B-Comp)}~\cite{FLMR04} (based on the above-mentioned idea by von Neumann):
\begin{enumerate}
  \item CNOT, spin $B$ as a control and spin $A$ as a target.\newline
        Spin $A$ is flipped if $B=1$.
  \item NOT on spin $A$.
  \item CSWAP, spin $A$ as a control, and spins $B$ and $C$ as
targets.\newline
        $B$ and $C$ are swapped if $A=1$.
\end{enumerate}
When the three spins have initial bias $\eps$, this  
subroutine cools spin $C$: if $A=1$ after the first
step (and $A=0$ after the second step),
$C$ is left unchanged (with its original bias $\eps$);
however, if $A=0$ after the first step ($A=1$ after the second step),
spin $B$ is cooled by a factor of about 2. The CSWAP places the new bias
on $C$, which is therefore, on average, cooled by a factor of $3/2$
(assuming biases much smaller than 1).
We do not care about the biases of the other two spins, as they will undergo reset. 
One may consider the case in which those two remaining spins have reset spins as
their neighbours.   
As a result, a single application of 3B-Comp, followed by reset of the two
(potentially heated) spins by a simple PT from their neighboring reset spins cools the
entire 5-spin system, once the reset spins go back to their initial equilibrium
temperature.

%
%
The following details are provided here to clarify the theoretical differences
between AC and polarization compression. We assume here that we are given 
a system of $n$ spins, all with the same polarization bias,
as in the case analyzed near the end of the previous section.
In contrast to the case analyzed earlier, now one of the $n$ spins is assumed to
be a 
reset spin, and the remaining $n-1$ spins are 
computing spins.
A PAC variant in which reset spins also participate in compression
is called PAC2 in~\cite{FLMR04}. PAC2 is simple - all 3-bit compressions are 
always applied to spins with
identical biases. 
On a 3-spin system (assuming the right-most is the reset spin), 
one gets the final biases $\{3/2, 1, 1\}$, in units of $\eps$. 
On a 5-spin system (again, assuming the right-most is the reset spin), 
one gets the final biases $\{9/4, 3/2, 3/2, 1, 1\}$, where the 
$9/4 = (3/2)^2$ bias is obtained by applying 3B-Comp onto 3 spin already cooled
to biases of $3/2$.
Generalizing this process to $n$ spins 
one gets 
(ideally) biases of 
$\{(3/2)^{(n-1)/2}, \ldots ,(3/2)^2, (3/2)^2, (3/2), (3/2), 1, 1\}$, in units of
 $\eps$, for any odd number of spins (out of which, $n-1$ are computing spins
and one is a reset spin).
PAC2 proves an exponential advantage of AC over the best reversible cooling,   
as the latter can only improve the bias of the coldest
 spin by a factor of $\sqrt{n}$. In typical scenarios, reset spins have higher
initial polarizations than computation spins; the bias amplification factor 
of $(3/2)^{(n-1)/2}$ is relative to the larger bias of the reset spin.

%
%
The cooling steps (reset and reversible polarization compression) 
can be repeated several times. 
Fernandez~\cite{Jose-PhD-Thesis} considered two computation spins and one 
reset spin and analyzed \emph{optimal cooling} of this system;
By repeating the reset and compression exhaustively, the final biases of the
 three spins approach the limit of $\{2,1,1\}$ in units of $\eps$, the
equilibrium bias of the reset spin.
Optimal AC~\cite{SMW05,EFMW06} 
leads to the  exponential series:
$\{... 128, 64, 32, 16, 8, 4, 2, 1, 1\}$, 
so the coldest spin is cooled by a factor of $2^{n-2}$. 

\subsection{Theoretical heat-bath cooling and algorithmic cooling of TCE}~\label{app:theoretical-HBC-and-AC-of-TCE}

TCE can be ideally considered as a 3-spin system, with one reset spin and two
computation spins. Its initial biases are $\{1, 1, 4\}$, with an information
 content (IC) of $18$. Bypassing Shannon's entropy bound means increasing the IC
of the 3-spin system above $18$.

We show in the main text 
that an ideal POTENT leads to $\{4, 4, 4\}$, with an information content of 
$48$, bypassing Shannon's entropy bound. 
A partial process, with just one selective reset, leads to
the biases $\{1, 4, 4\}$
(or the biases $\{4, 1, 4\}$),  
with an information content of 33, 
which already bypasses 
Shannon's entropy bound. 
Note that this ``single selective reset'' contains only one PT and one reset
(or a dual PT in case of transferring the polarization to the far carbon).
Still, we did not find any paper reporting a bypass of the entropy bound via this process. 
POTENT, of course, is more powerful, as the repeated selective reset cools
two computing spins, and can also lead further, to AC\@.

Some interesting (ideal) cooling processes with TCE are described below.
By one selective reset step to the far carbon, plus one PT from the proton
to the near carbon, the resulting biases become 
$\{4, 4, 0\}$ in the case of a PT (or
$\{4, 4, 1\}$ in case of an ideal SWAP)
with an information content of $32$ ($33$).  
This process is important, since it focuses on cooling the two spins of interest,
such that the system is cooled beyond the entropy bound. 
Our ideal simulation (see section~\ref{sec:simulations}) shows that it may be 
achieved given the relaxation times in our system. However, a practical simulation, 
accounting for realistic PT steps (see again, section~\ref{sec:simulations}), 
shows that our current experimental parameters
do not support it, and more efficient PT steps are required. 
Our experimental efforts yielded cold carbons (see main text and
 Appendix~\ref{sec:exp-add}),
but no bypass of the entropy bound in this interesting case; however, see the
Post Scriptum for some very recent results.

Full AC can cool further.
As explained above, an ideal PAC
cools the far-carbon to $4 \times 3/2 = 6$, 
which bypasses the entropy bound already on a \emph{single} spin, as 
bypassing $\sqrt{18}$ on that cooled spin is sufficient for 
that purpose. In the ideal case (biases of $\{4, 4, 4\}$), 
a perfect compression
followed by resets of the two remaining spins 
would yield approximately (error in $\eps^3$) $\{6, 4, 4\}$. The information
content of this final state is $68$, bypassing, by far, 
the entropy bound. 

Interestingly, a perfect optimal compression applied onto 
biases of $\{4, 4, 4\}$ leads immediately to biases of 
$\{6, 2, 2\}$  
and information 
content of $44$, which is lower than 
the information content of the initial state (48) in that case; The residual information content
is transfered into classical correlations (compression is unitary and uses classical
gates, so the total entropy cannot change by that process).
Our practical simulation shows that bypassing~4 on the colder carbon is not
possible with our current system (see section~\ref{sec:simulations}).

A similar analysis, by the way, is relevant for understanding the differences
between PT and SWAP: both are unitary and preserve
entropy, but PT (e.g. INEPT) moves some of the information content into classical
correlations, 
so that (for two spins) initial biases of $\{4, 1\}$ change to 
final biases of $\{0, 4\}$ by ideal PT, while they change to 
final biases of $\{1, 4\}$ by ideal SWAP.

The optimal cooling of a single spin leads to biases of $\{8, 0, 0\}$ and an information content of 64.
The optimal cooling of the entire 3-spin system leads to biases of 
$\{8, 4, 4\}$ and an information content of 96.
Both require many reset steps and are thus rather impractical due to decoherence and
imperfect PT and compression steps; however, 
even reaching biases around
$\{5, 3, 3\}$ and an information content of 43 could be very interesting.
See the Post Scriptum for some very recent results approaching such goals.
 
\section{The experiments}\label{app:the-experiment}

\subsection{Experimental details}\label{sec:exp-det}
%
%
The initial polarization biases were calculated according to the high
temperature approximation, $\eps \approx \Delta E/2k_BT$ (see
 Appendix~\ref{app:effective-cooling-approaches-in-NMR}), 
where $\Delta E = h\nu$, $h$ and $k_B$ are the Planck and Boltzmann constants, and
 T is the room temperature of 296K$ \pm 1$K. The resonance frequencies, $\nu$, 
 determined by averaging the resonance frequencies of the four lines of the
multiplet of each spin, were $\nu^\h = 500.1332346\cdot10^6 \pm 0.9\ \Hz$,
 $\nu^\Cone = 125.7735301\cdot10^6 \pm 0.7\ \Hz$, and
 $\nu^\Ctwo = 125.7726234\cdot10^6 \pm 0.4\ \Hz.$
The line-width of all lines was below $2\ \Hz$ ($1.8\ \Hz,$ $1.4\ \Hz,$ and
$0.8\ \Hz$ for \h, \Cone, and \Ctwo, respectively). 
The resulting biases were $\eps^\h = 4.05\pm0.01\cdot10^{-5}$ and
 $\eps^\Cone = \eps^\Ctwo = 1.020\pm0.003\cdot10^{-5};$ the error due to the
high temperature approximation ($\eps \approx \tanh(\eps)$) was much smaller, in
 the ninth and tenth decimal place for the proton and carbons, respectively. 
The biases were normalized by
 setting the bias of \Ctwo, which was on resonance, to $\epsn^\Ctwo = 1.000$ 
with $3\%$ error due to the uncertainty in room temperature.
The error values in the initial biases were derived here 
by the standard 
error formula:
$$
\Delta \epsn = \sqrt{\left(\frac{\partial \epsn}{\partial \nu}\Delta
 \nu\right)^2+\left(\frac{\partial \epsn}{\partial T}\Delta T\right)^2} =
 \epsn\left[\sqrt{\left(\frac{\Delta \nu}{\nu}\right)^2+
 \left(\frac{\Delta T}{T}\right)^2}\right].
$$

%
%
A block diagram depicting the various stages of our experiment is
shown in Figure~\ref{fig:full-experiment}. The first and third transfer
sequences (both from \h\ to \Ctwo) use different refocusing schemes;
the latter was designed to retain the enhanced polarization of \Cone\ by
 refocusing its evolutions (indicated in Figure~\ref{fig:full-experiment} by
 $*$).
The first two polarization transfers are implemented by two overlapping
refocused-INEPT\footnote{The term refocused-INEPT coined by Burum and
 Ernst~\cite{BE80} is not connected to the term refocusing used elsewhere
in this paper with respect to canceling out undesired evolutions.}
 sequences~\cite{BE80}, as shown in Figure~\ref{fig:pulseq-annot}a and
 Figure~\ref{fig:pulseq-annot}b.
After the second refocused-INEPT, a $90^\circ$ pulse aligns the carbons along
the z axis prior to the first WAIT (see Fig.~\ref{fig:pulseq-annot}c).
The third refocused-INEPT is immediately followed by a selective inversion 
of~\Cone, and prior to the second WAIT, the carbons are aligned along the z axis
by a $90^\circ$ pulse (see Fig.~\ref{fig:pulseq-annot}d). Finally, acquisition 
is performed measuring the free induction decay along the y 
axis\footnote{Equivalent to taking the trace of 
$\left(I^+\equiv I_x+iI_y\right)\rho,$ where $\rho$ is the final density matrix.} 
for either the carbons (shown in Fig.~\ref{fig:pulseq-annot}d), 
or the proton. During the second WAIT, the spins
 are aligned along the z axis, hence the need for a final $90^\circ$ pulse, see
 Fig.~\ref{fig:pulse-sequence}.

%
%
Recall that a transfer of polarization can be bi-directional
(achieved by implementing a SWAP gate),
yet we chose to use a uni-directional PT, 
so that the resulting pulse sequence is more efficient. 
We observed that 
each INEPT-based PT
sequences implements (up to some irrelevant phases) 
a dual-CNOT gate, namely 
CNOT(source,target)-CNOT(target,source),
which is preferred here over a SWAP gate,
that is equivalent to a triple-CNOT gate, 
namely 
CNOT(source,target)-CNOT(target,source)-CNOT(source,target).
The PT pulse sequence is preferred over the SWAP 
pulse sequence since it is shorter and
contains fewer basic pulses, say, about two-thirds. 

%
%
The complete POTENT sequence, detailed in Figure~\ref{fig:pulse-sequence},
was run with various combinations of $t_1$ and $t_2$ delays.  For statistics,
several spectra were acquired under the same conditions. Reported values 
were obtained in five single-scan measurements (for each nucleus).
In order to validate the sequence and estimate the transfer efficiencies,
truncated versions of the complete pulse sequence were acquired, each version
terminating at a different stage. Intermediate spectra obtained in this manner 
are shown in Figure~\ref{fig:spectra}.

%
%
We obtained efficiencies of $92\% \pm 2\%, 69\% \pm 1\%,$ and $74\% \pm 1\%$, 
for the three PTs. These transfer efficiencies were obtained by comparing the
peak integrals after truncated pulse sequences at the laboratory conditions of 
the full experiment (errors are due to uncertainty in the integrals). 

\subsection{Additional experimental results: cooling two spins}
\label{sec:exp-add}
%
%
It would be of interest if the combined entropy of the carbons after 
truncated 
POTENT, that excludes the final WAIT step, 
 would exceed the equilibrium entropy of the spin system. This would occur, for
instance, if both biases increased three-fold, thereby cooling both carbons to
 one third of the room temperature. To ascertain whether this could be achieved,
 we performed experiments with the POTENT sequence with a very short second
 delay, about 2--3$\Ttwo^*,$ required for the elimination of coherences that
 cause undesired phases\footnote{$\Ttwo^*$ denotes the apparent \Ttwo,
reflected in our experiments by the linewidth.}. 
By setting the second delay, $t_2$, to $0.4\sec$, the
carbon re-heating is minimized. We reached biases of 
$2.03 \pm 0.02$ and $2.91 \pm 0.04$ for \Cone\ and \Ctwo, respectively\footnote{
For \Cone, the error was obtained by linear regression over 8 results obtained
 with $d2=1,2,\ldots,8\sec$ ($R^2=0.999$), while for \Ctwo the error was
 obtained by logarithmic regression over 5 data points ($d2=4,5,\ldots,8\sec,$
$R^2=0.995$).}, yielding $I(\Cone,\Ctwo)=12.6 \pm 0.2$.
Although we did not succeed to bypass the entropy bound in this case, the two
carbon spins were cooled quite significantly. 
These biases  were obtained with 
 $t_1 = 8\sec$, and the resulting
 temperatures were $145 \pm 2$K  and $101 \pm 1$K for \Cone\ and \Ctwo, respectively,
 as mentioned in the ``Results'' section of the main paper. 

%
%
In experiments designed to achieve ``elementary bypasses'',
namely, in ``single-selective-reset'' experiments, 
the combined information contents of the proton and the 
cooled carbon 
were sufficient to bypass the entropy bound.
For \Cone\ and \h, we obtained biases of $1.61$ and $3.95,$
with $t_1 = 2\sec$ and $t_2 = 18\sec,$ yielding 
 $I(\Cone,\h)=18.17.$ 
[In another experiment we obtained, 
for \Cone\ and \h, biases of $1.64$ and $3.94,$
with $t_1 = 4\sec$ and $t_2 = 18\sec,$ yielding (again) 
$I(\Cone,\h)=18.17.$] 
For \Ctwo\ and \h, we obtained biases of $1.90$ and $3.80,$
with $t_1 = 12\sec$ and $t_2 = 12\sec,$ yielding 
 $I(\Ctwo,\h)=18.03.$ 
[In another experiment we obtained, 
for \Ctwo\ and \h, biases of $1.79$ and $3.85,$
with $t_1 = 12\sec$ and $t_2 = 14\sec,$ yielding (again) 
 $I(\Ctwo,\h)=18.03.$]
These results were obtained with no optimization, however they are 
already beyond the entropy bound, when considering the $\pm 0.1$ error in 
information content obtained over five repetitions in the main experiment.

\section{Post-Scriptum details}\label{app:post}
%
%
\paragraph{Accounting for longitudinal relaxation.}
The limitation of algorithmic cooling due to finite ratios, \RR, between
 the \Tone\ relaxation times of computation spins and the characteristic
repolarization times $\tau$ of reset spins are discussed in
Refs~\cite{EMW11,BEMW}. The duration of each reset step is
$\Twait = d\cdot\tau,$ where in liquid state NMR $\tau$ is \Toner, while in
the solid state~\cite{BMR+05,RMBL08} $\tau$ was the characteristic time for
 spin diffusion.
In Ref~\cite{BEMW} we analyze various cooling algorithms for several values
of \RR. 
Ideally, $\Tonec \gg \Trun \gg \Twait \gg \tau,$ where $\Trun =
 \Twait\cdot\NRes$ is the runtime of the algorithm and \NRes is the number of 
reset steps.
A partial analysis is given in Ref~\cite{EMW11} for 2PAC. PAC2~\cite{FLMR04}
may ideally cool one spin in a 5-spin system that includes one reset spin
by a factor of $2.25$ after 9 reset steps~\cite{FLMR04,EFMW06}.
When $\RR = 10000$ (similar to \RR in Refs~\cite{BMR+05,RMBL08}) and $d=5,$
 the number of reset steps, $d\cdot\NRes = 5\cdot17 \ll \RR,$ and the cooling
 factor is essentially unchanged ($2.23$). However, when $\RR \sim d\cdot\NRes,$
the deviations become significant, e.g. the cooling factor reduces to 2.14
 (1.81; 1.76) for $\RR = 100 (10; 5).$ In such cases, better cooling is obtained by choosing lower values of $d$; for example, when $\RR = 10$ and $d=3,$ the
cooling factor is improved to $1.85.$ 
  
%
%
\paragraph{Heat-bath cooling of amino acid spin systems.}
In Ref~\cite{EGMW11}, we applied heat-bath cooling to the \isotope{C}{13}
labeled backbone of two amino acids, glutamate and glycine. Unlike TCE, the
large chemical shift between the two labeled carbons allowed highly efficient
1 \msec selective pulses for excitation and inversion of either carbon. 
Sufficient \Tone\ ratios ($\RR \sim 10$) were found between \Cone\ and the alpha
 proton(s), However, the other \Tone\ ratio (involving \Ctwo) was quite low
 (around 2). Therefore, in order to reduce the total entropy of the spin system
 (beyond Shannon's bound) we applied a single selective reset such that only
 \Cone remained cool. In addition, we applied truncated POTENT sequences, where
 the final delay was omitted, to significantly cool both carbons (each by a
 factor of about $2.5$). Similar results were obtained for glutamate under 
 physiological conditions (temperature and pH).

%
%
\paragraph{Compression and AC of TCE.}
We attempted to perform the polarization compression step on TCE using
non-selective pulses, as the small chemical shift did not permit efficient
 spin-selective pulses\footnote{An earlier 3-bit compression
 experiment~\cite{CVS01} employed spin-selective pulses and had very long
 coherence times (several seconds), yet the efficiency was still quite low (about 50\%).}. 
 However, significant enhancement could not be obtained,
due to accumulation of errors and extensive decoherence over the many (dozens)
of pulses and delays. Highly efficient 3-bit compression was achieved in the
 solid state using quantum optimal control theory
 (the GRAPE algorithm)~\cite{BMR+05,RMBL08}. 
We have recently~\cite{Atia-MSc-Thesis,AEMW-1,AEMW-2}
 adopted this approach in the liquid state, and achieved short optimized
 3-bit compression shaped pulses (around 15 \msec) for TCE; the high efficiency
 of the compression ($\sim 90\%$) allowed us to cool one carbon of TCE by about
 4.5-fold, beyond Shannon's bound, following several rounds of algorithmic
 cooling. 

\end{document}